\documentclass{article}

\AtBeginDocument{%
  }

\usepackage[accepted]{mlsys2025}

\usepackage{microtype}
\usepackage{graphicx}
\usepackage{booktabs}
\usepackage{enumitem}
\usepackage{wrapfig}
\usepackage{xspace}
\usepackage{datetime}
\usepackage{censor}

\usepackage[colorinlistoftodos,disable]{todonotes}
\usepackage{hyperref}

\usepackage{amsmath}
\usepackage{amssymb}
\usepackage{mathtools}
\usepackage{amsthm}
\usepackage{cleveref}

\theoremstyle{plain}

\theoremstyle{definition}

\theoremstyle{remark}

\usepackage{tikz}
\usepackage{algorithm}
\usepackage{algorithmic}
\usepackage{subcaption}
\usepackage{multirow}

\newcommand{\INPUT}{\item[\textbf{input}]}

\newcommand{\google}{{Google\xspace}}

\begin{document}

\twocolumn[
\mlsystitle{A Bring-Your-Own-Model Approach for ML-Driven Storage Placement in Warehouse-Scale Computers}

\mlsyssetsymbol{equal}{*}

\begin{mlsysauthorlist}
\mlsysauthor{Chenxi Yang}{ut}
\mlsysauthor{Yan Li}{goo}
\mlsysauthor{Martin Maas}{gdm}
\mlsysauthor{Mustafa Uysal}{goo}
\mlsysauthor{Ubaid Ullah Hafeez}{goo}
\mlsysauthor{Arif Merchant}{goo}
\mlsysauthor{Richard McDougall}{goo}
\end{mlsysauthorlist}

\mlsysaffiliation{ut}{UT Austin. Work was done while at Google.}
\mlsysaffiliation{goo}{Google}
\mlsysaffiliation{gdm}{Google DeepMind}

\mlsyscorrespondingauthor{Yan Li}{elliotli@google.com}

\mlsyskeywords{Machine Learning for Systems, Data Placement Optimization, Data Centers, Storage Systems}

\vskip 0.3in

\begin{abstract}
Storage systems account for a major portion of the total cost of ownership (TCO) of warehouse-scale computers, and thus have a major impact on the overall system's efficiency. Machine learning (ML)-based methods for solving key problems in storage system efficiency, such as data placement, have shown significant promise. However, there are few known practical deployments of such methods. Studying this problem in the context of real-world hyperscale data centers at \google, we identify a number of challenges that we believe cause this lack of practical adoption. Specifically, prior work assumes a monolithic model that resides entirely within the storage layer, an unrealistic assumption in real-world deployments with frequently changing workloads.
To address this problem, we introduce a cross-layer approach where workloads instead “bring their own model”. This strategy moves ML out of the storage system and instead allows each workload to train its own lightweight model at the application layer, capturing the workload's specific characteristics. These small, interpretable models generate predictions that guide a co-designed scheduling heuristic at the storage layer, enabling adaptation to diverse online environments. We build a proof-of-concept of this approach in a production distributed computation framework at \google. Evaluations in a test deployment and large-scale simulation studies using production traces show improvements of as much as 3.47$\times$ in TCO savings compared to state of the art baselines. 

\end{abstract}
]

\printAffiliationsAndNotice{}

\section{Introduction}
\label{sec:intro}

Storage systems comprise a large part of data centers' total cost of ownership (TCO). Even small improvements in storage system efficiency can have a major impact on the overall costs. Improvement as low as 1\% represents a large amount in the context of hyperscale data centers, which see billions of dollars of investment \cite{dataCenterInvestment}. Data placement – e.g., deciding whether a file should be stored on hard disk (HDD) or solid state drives (SSD) – is an important decision impacting the efficiency and costs in data center storage systems. This problem is also known as storage tiering and has been the subject of a large amount of research \cite{kim2014evaluating, dulloor2016data, saxena2012flashtier}.  Currently, there are two approaches to this problem:

\begin{itemize}[itemsep=0.5pt,wide = 0pt,topsep=0.2pt]
    \item \textbf{Heuristics}, such as greedily allocating data to SSDs until capacity is reached and using HDDs for overflow \cite{albrecht2013janus, yang2023fifo, eytan2020s, yang2013hec, yang2022cachesack}. These heuristics are deployed and represent today's state-of-the-art. They are fast and interpretable, but perform suboptimally when SSD capacity is limited.
    \item \textbf{Machine Learning (ML)} approaches that leverage real-world workload information \cite{zhou2021learning}. Few of these approaches are practically deployed due to considerations such as run-time overhead, adaptability, or decomposition challenges, and risks associated with a model becoming a single point of failure  \cite{maas2020taxonomy, paleyes2022challenges}.
\end{itemize}

\begin{figure}
    \centering
    \includegraphics[width=\linewidth]{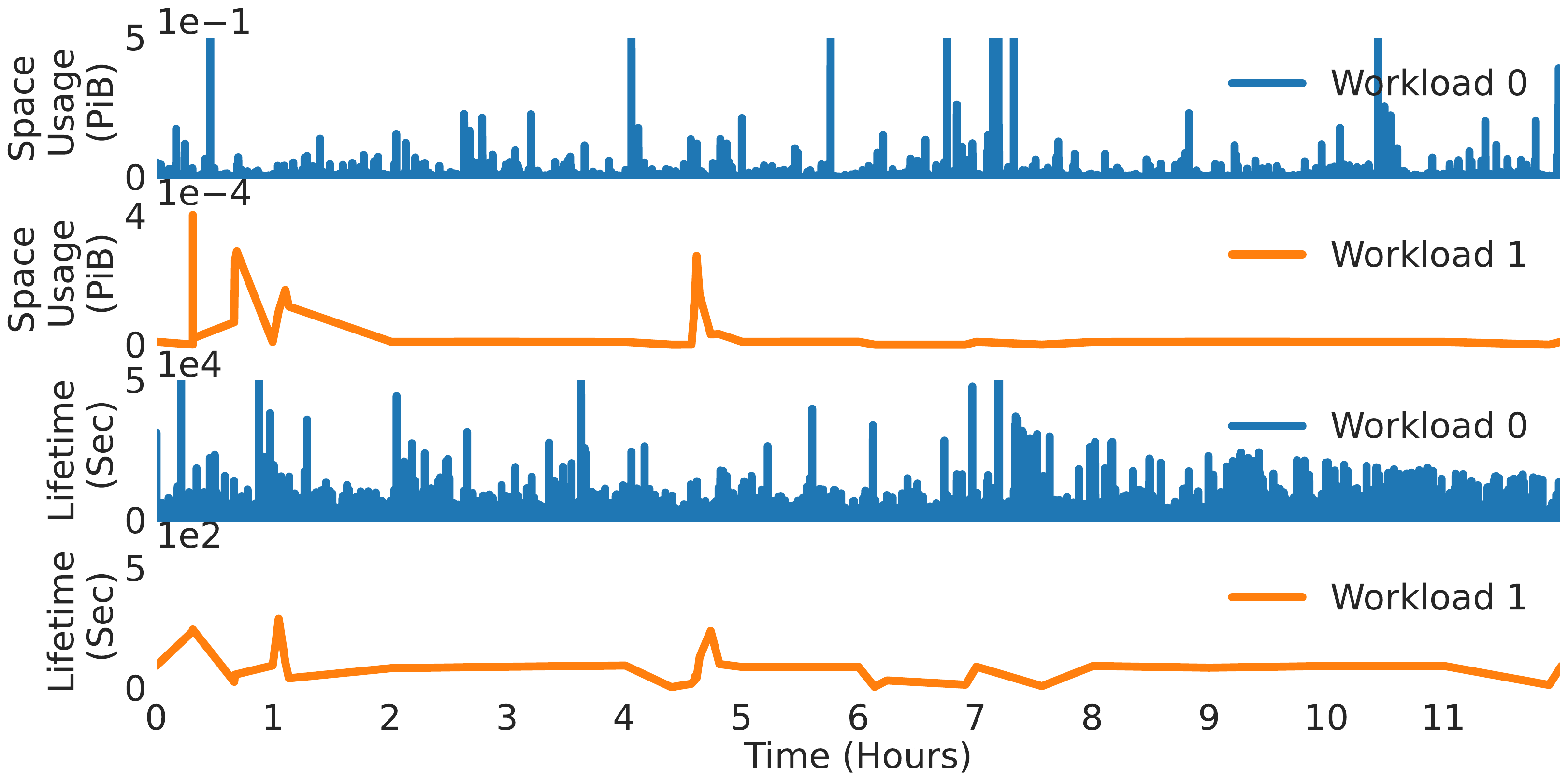}
    \caption{Workloads show vastly different storage patterns.}
    \label{fig:workload_pattern}
\end{figure}

\noindent To understand the challenges behind adoption of ML in real-world scenarios, we analyzed our real-world production systems at \google. Data centers run a wide range of workloads with vastly different characteristics (\Cref{fig:workload_pattern}). Workloads arrive and evolve at a high rate, and data access patterns are highly dynamic and application-specific. Data centers deal with this issue through \emph{abstraction layers}, such as the application, storage or hardware layer: For example, the application layer does not need to worry about the specifics of the hardware, and the storage layer does not need to know any details about the inner workings of each application. This enforces separation of concerns and allows these components to operate and evolve independently.

We posit that this is the key challenge behind existing proposed ML methods. Existing ML works mostly treat end-to-end data placement as one problem and assume a monolithic model deployed within the storage system \cite{liu2020imitation, singh2022sibyl, zhou2021learning, kaler2023deep}. Such a model might be trained on file names or common application behavior \cite{zhou2021learning}. While this approach works in simulation, it breaks the separation of concerns, which is problematic in real-world large-scale systems. For instance, changes to a major application that affect file names would need to trigger a retraining of the model at the storage layer, which needs to roll out changes at a much lower velocity than the workloads.

\begin{figure}
    \centering
    \includegraphics[width=\linewidth]{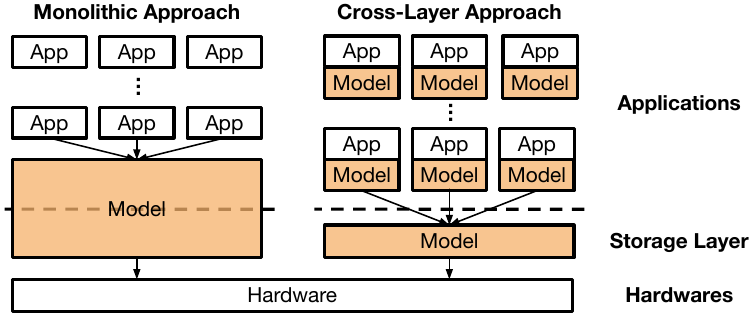}
    \caption{Conceptual overview of the monolithic approach vs. the cross-layer approach.}
    \label{fig:mono-vs-cross}
\end{figure}

We introduce a ``Bring-Your-Own-Model'' (BYOM) design that embraces the multi-layer characteristic of storage systems and presents a practical solution to these problems by combining cheap and interpretable ML models at the application layer with a custom algorithm that leverages their predictions at the storage layer. Instead of a single large ML model, we build smaller models for individual workloads, which produce hints that the storage layer can utilize to place the workloads' data (Figure~\ref{fig:mono-vs-cross}).

Evaluating this approach at scale presents a chicken-and-egg problem. Porting a sufficiently large number of complex workloads to this approach is a large investment and requires the approach to be in place, but putting the approach in place requires experimental validation with a large range of workloads. We side-step this issue by focusing on a particular class of workloads that are written in a distributed data processing framework similar to Apache Beam \cite{apachebeam2012} or Apache Spark \cite{zaharia2012resilient}. This framework is used for a wide range of workloads in our fleet, including log processing, simulations, streaming applications, and a variety of ML workloads (which may use accelerators). By targeting tiering of intermediate files generated by these workloads, we can use large-scale historical traces and evaluate our approach across a highly diverse set of applications (\Cref{fig:workload_pattern}), without having to port each of them manually. Note that cumulatively, these files can account for a significant portion of all storage resources in a data center (up to 35\% in some clusters).

We first present a headroom analysis to understand the potential upside from ML over traditional heuristics. We formulate the data placement problem into an Integer Linear Programming (ILP) problem and use a solver to determine optimal placement decisions. We find that these optimal decisions can achieve 5.06$\times$ the cost savings of a state-of-the-art heuristic approach (but require clairvoyant knowledge).

Prior work has proposed imitation learning against such an oracle \cite{liu2020imitation}. However, we find that this approach does not work in our deployment, since the model does not only need to make decisions for individual workloads but \emph{adapt} to an environment that is changing due to external factors (e.g., varying load patterns and workloads arriving or leaving).
Our BYOM design tackles this \emph{adaptability problem}. At the application layer, we analyze the data properties that contribute to the optimal placement. We then design a model to predict a ranking based on these properties, which is independent of online fluctuations of the environment and other applications. Finally, we design an adaptive algorithm at the storage layer to select the data to place on SSD based on the model predictions from all applications and system feedback.

We instantiate this design in the context of a production data processing framework at \google. We show its practicality by developing a prototype of our approach and running it in a test deployment. We also perform an extensive simulation study based on real-world production traces from \google's data centers. We show that our approach can lead to an additional 3.22\% TCO savings, more than 3.47$\times$ the savings from the production baseline.
We summarize our contributions as follows:

\begin{itemize}[itemsep=0.5pt,wide = 0pt,topsep=0.2pt]
    \item Within a real-world production setup, we investigate ML for storage placement from a new perspective, with a focus on practicality in production settings.
    \item We design and instantiate a novel cross-layer approach, combining ML and heuristics that can adapt across workloads and external factors in data centers.
    \item We prototype the proposed ML integration to show its realizability in a real production codebase.
    \item We evaluate our method at scale with production traces and achieve $3.47\times$ TCO savings over SOTA baselines.
\end{itemize}

We first present background on storage systems and production constraints (\Cref{sec:bg}), and formulate our optimization problem with baselines (\Cref{sec:prob}). We then discuss our approach (\Cref{sec:sys}), introducing our ML method and scheduling algorithm. We next show detailed prototype and large-scale simulation studies (Section~\ref{sec:eval}). Finally, we discuss related work (Section~\ref{sec:related-work}) and conclude (Section~\ref{sec:conclude}).
\section{Background}
\label{sec:bg}

\subsection{Storage for Data Processing Frameworks}
\label{sec:storage-for-data-processing-frameworks}

\begin{figure*}[t]
    \centering
    \includegraphics[width=0.65\linewidth]{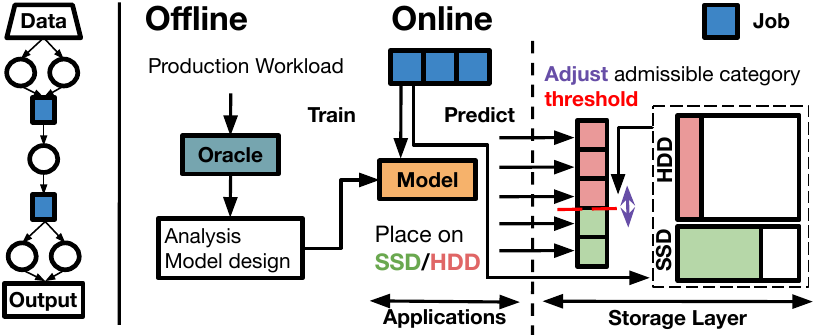}
    \caption{\textbf{Left:} Data flow graph in a data processing framework. Data is processed in parallel and its jobs create intermediate files (blue) which are inputs for the next processing step. \textbf{Right:} Approach Overview. We analyze production workloads offline for model design and training. Online, each application's model predicts job properties and passes the prediction to the storage layer for job placement.}
    \label{fig:overview}
\end{figure*}

Modern data processing frameworks, such as Apache Beam, structure their computations as data flow graphs (\Cref{fig:overview}, left). Each node (or \emph{step}) within the graph represents a computation step. Edges represent the flow of data. Computations are highly parallel, and a distributed framework spawns \textit{workers} to execute tasks. A worker is a process that runs on a server. Data is generally passed between workers through \textit{shuffle jobs}. A shuffle job is generated when the execution of the workflow reaches a step or operation that necessitates the exchange of information. As an example, \emph{GroupByKey} is a common operation across frameworks that generates one or more shuffle jobs. During a shuffle job, workers write their data as \emph{intermediate files} to a distributed file system \cite{hdfs2010, calder2011windows, ghemawat2003google} and read it in subsequent steps. The access patterns to the files depend on the specifics of the computation, such as filtering, grouping, or sorting. One shuffle job can operate on multiple intermediate files.

\subsection{SSD/HDD Tiering and its Trade-Offs}

Storage cost falls into several different categories: 1) the amount of storage (e.g., in GiB) occupied by the data, 2) wearout of devices such as SSD, which degrade with every write, and 3) the amount of I/O operations (i.e., read or write requests to a storage device per unit time sustained by the device). SSDs and HDDs have different trade-offs among all three dimensions. SSDs provide much larger amounts of I/O with a higher cost per GiB and write-induced wearout. In contrast, HDDs are ideal for large amounts of data and long sequential access patterns that introduce few I/O operations. At the same time, SSDs are ideal for random, small accesses – if the resulting wearout can be tolerated. In practice, intermediate files in data processing pipelines can fall into either category (or, more commonly, in between), which makes the data placement problem challenging.

\subsection{Production Requirements and Limitations}
\label{ssec:production}

Data centers accommodate a vast array of workloads with diverse behavior patterns. Employing a single model in the storage layer to jointly learn all workloads introduces a single point of failure, and requires the model to be large and complex, and thus expensive and difficult to interpret.
Further, this approach requires all input features to be reliably delivered to the storage system, when some of the most predictive features are workload-specific \cite{zhou2021learning}. Finally, In hyperscale data centers, workloads exhibit significantly faster rates of change than the update cycles of storage systems. This causes a dilemma: 1) Rolling out the model with the storage system means it is stale by the time it reaches production; 2) updating the model independently of the storage system means that it is not tested as rigorously as the rest of the system, thus becoming its weakest link.

To address these problems, we propose a more granular approach where each workload has its own dedicated model to produce a \emph{hint}, which is then reliably passed to the storage system. This hint indicates, for example, how well a file can be cached. Since workloads ``\emph{bring their own model}'', models evolve at the velocity of the workload rather than the storage system. Because the models are smaller, they are cheaper and more interpretable.
They are distributed across many workloads hence they can use more features and are more robust: a model failure only affects one workload. In this work, we focus on optimizing data placement for workloads built on a particular data processing framework, but our cross-layer design is general.

Training a model offline and deploying it online is challenging, since a static model cannot adapt to evolving workload patterns, which are prevalent in real-world scenarios. To address this issue, we present an adaptive strategy that utilizes online observations to inform placement decisions.

Models can introduce non-trivial overheads. For example, prior work suggested using Transformers, which are known for their superior learning capabilities but can be computationally expensive, incurring prediction latency costs of approximately 99ms \cite{zhou2021learning}. To balance performance and efficiency, we leverage gradient boosted trees as the model, providing a compromise between lightweight, low-performance models and powerful, expensive models. 

\subsection{\google's Production Setup}

Our fleet consists of clusters of up to O(10,000) machines. Dedicated storage servers handle all requests to the distributed storage system. SSD and HDD devices are hosted by different servers due to different device form factors. While regular servers can also have locally attached storage, it is mainly used by the local OS and not for compute data.

A global distributed file system is used to store files. The system provides a client library that resides on all compute servers and communicates with storage servers. Data processing frameworks running on top of compute servers use this library to store and retrieve files. SSD tiering is handled by a service running on a dedicated set of servers that analyze the client traffic and make caching decisions (admit to/delete from SSD, evict to HDD,...). They proxy requests to the appropriate storage servers. This work focuses on the SSD/HDD tiering decisions within these caching servers. More details are shown in \Cref{app:production-details}.
\section{Problem Formulation \& Baselines}
\label{sec:prob}

We now provide a concrete definition of the data placement problem. Our basic data placement unit is a shuffle job. We track four attributes for each job: (start time, lifetime, job size, cost). Job size is measured in bytes. A cluster has an SSD capacity. The placement algorithm produces a mapping: $\text{job} \rightarrow \{\text{SSD}, \text{HDD}\}$.

Resource savings from SSD tiering are relative to a baseline where all files are stored on HDD. Our main optimization criterion is storage resources. While SSD tiering may also reduce CPU/RAM resource consumption of individual jobs and improve their tail latency, those improvements are \emph{opportunistic} and not the target of our work. Since workloads are written assuming HDD storage, they do not rely on these speed-ups. Further, we found that \emph{aggregate} CPU/RAM resource consumption across all jobs is not significantly affected by the tiering technique that is used.

We evaluate resource savings from two perspectives. First, we measure the reduction in HDD I/O from moving jobs to SSDs. This frees up HDD I/O for operations that cannot (or should not) be moved off HDDs, such as accesses to cold data. We quantify this reduction using a metric called \emph{Total Cost of I/O} (\textsc{TCIO}), where a value of 1.0 represents the amount of I/O that a standard HDD can sustain per second. Jobs running on SSDs have a \textsc{TCIO} of zero. Our TCIO calculation accounts for caching effects, such as the DRAM cache present alongside HDDs in each server. In our system, I/Os that are served from cache do not reach the disks, and small write operations are grouped into 1 MiB chunks before reaching the disks. This ensures that TCIO reflects the true workload pressure on the disks.

Second, we measure overall monetary savings. We define \emph{Storage Total Cost of Ownership} (TCO) as the total expenditure associated with acquiring, operating, and maintaining a storage system. TCO is calculated separately for HDDs and SSDs due to the different nature of these devices. Substitute $\textsc{Dev}$ for HDD or SSD below to get the respective definition:

\begin{equation*}
\begin{aligned}
\mathrm{TCO^{\textsc{Dev}}} &=  \mathrm{cost^{\textsc{Dev}}_{byte}} +
\mathrm{cost^{\textsc{Dev}}_{network}} + \mathrm{cost^{\textsc{Dev}}_{server}} \\
    &\phantom{{}={}} + \mathrm{cost^{\textsc{Dev}}_{specific}} \\
\mathrm{cost^{\textsc{Dev}}_{byte}} &=  \mathrm{byte\_cost^{\textsc{Dev}}} \cdot \mathrm{size} \cdot \mathrm{duration}\\
\mathrm{cost^{\textsc{Dev}}_{network}} &= \mathrm{network\_cost\_rate} \cdot \mathrm{IO\_throughput^{\textsc{Dev}}} \\
  &\phantom{{}={}} \cdot \mathrm{duration} \\
\mathrm{cost^{HDD}_{server}} &= \mathrm{server\_cost\_rate^{HDD}} \cdot \mathrm{\textsc{TCIO}} \cdot \mathrm{duration}\\
\mathrm{cost^{SSD}_{server}} &= \mathrm{server\_cost\_rate^{SSD}} \cdot \mathrm{IO\_throughput^{SSD}}\\
\mathrm{cost^{HDD}_{specific}} &= \mathrm{device\_cost\_rate^{HDD}} \cdot \mathrm{\textsc{TCIO}} \cdot \mathrm{duration} \\
\mathrm{cost^{SSD}_{specific}} &= \mathrm{wearout\_cost\_rate^{SSD}} \cdot \mathrm{total\_written\_bytes} \\
\end{aligned}
\end{equation*}

where $*\_\mathrm{cost\_rate}$ denotes conversion rates to dollar cost; $\mathrm{cost^{\textsc{Dev}}_{byte}}$ denotes the cost of storing one byte for one second on a device; $\textsc{TCIO}$, size, and duration denote a job's $\textsc{TCIO}$ need, storage footprint, and duration (for example, if a job has a TCIO of 2, the job would need two HDDs to run). The $\mathrm{cost^{\textsc{Dev}}_{network}}$ is a value derived from the data center total network cost of transmitting data at the aggregated throughput for the duration of the job. The network cost is largely constant and independent of the device, but we include it to avoid overestimating the impact of the other costs on the overall TCO. $\mathrm{cost^{HDD}_{server}}$ and $\mathrm{cost^{HDD}_{specific}}$ cover the cost of the servers and HDDs.

In practice, we found that the server cost for running a job on SSD correlates with the bytes transmitted. Because all SSD devices have a limit on the amount of Program/Erase (P/E) cycles and each write causes a loss in monetary value, the $\mathrm{cost^{SSD}_{specific}}$ is included to cover these wearout costs. $\mathrm{wearout\_cost\_rate^{SSD}}$ is calculated from the specific SSD drive model's total bytes written rating.

\subsection{Oracle: Optimal Solution Based on Solver.} To better understand the headroom that is available if we achieve perfect data placement, we design an oracle. It is an upper bound on the best solution, but impossible to implement. The oracle policy is based on using an Integer Linear Programming (ILP) solver and clairvoyant knowledge --- that is, assuming we know the future access pattern. We formulate the placement problem as an ILP problem:
\begin{equation*}
\begin{aligned}
&\max \sum_{i \in l} x_i (c^\text{HDD}_i - c^\text{SSD}_i)\\
\text{subject to:} \quad \quad &x_i \in \{0, 1\}, \forall i \in [0, N] \\
&p_i(t) = x_i s_i, \forall i \in [0, N], a_i \leq t \leq e_i \\
&\sum_{i \in [0, N], a_i \leq t \leq e_i} p_i(t) \leq M, \forall t \in T \\
&T = \max\{e_i : i \in [0, N]\}
\end{aligned}
\end{equation*}
The SSD space limit is $M$ and we assume the HDD space is infinite due to its lower cost per GiB. $X=\langle x_0,\dots,x_N\rangle$ is a sequence of arriving jobs. Job $i$, represented by variable $x_i$, arrives at time $a_i$, ends at time $e_i$, and it needs $s_i$ space with $c^\text{SSD}_i$ cost to put on SSD, $c^\text{HDD}_i$ cost to put on HDD. Oracle optimization can either optimize for \textsc{TCIO} or TCO. We use a binary variable, $x_i$, to denote the data placement decision, $x_i=1$ if $i$ is put on SSD, and $x_i=0$ if $i$ is put on HDD. Once placed, a job $x_i$ runs from $a_i$ to $e_i$ and $p_i(t) = x_i s_i$ represents the job's SSD consumption at time $t \in [a_i, e_i]$. Now the problem becomes maximizing the savings by placing jobs on SSD under space constraints (SSD space is limited): We run the above ILP with historical production workload data and find the optimal placement decisions that save the maximum amount of cost. The solver is optimal because it has clairvoyant knowledge. The clairvoyant knowledge includes information that is not available at a job’s placement decision time in practice: 1) The solver knows the global job ranking in terms of cost savings and would prioritize putting high cost saving jobs onto SSD. 2) The solver knows the workload patterns and places jobs that would monopolize SSD resources on HDD instead.

In addition to clairvoyant knowledge, the oracle needs a fixed SSD capacity limit for optimal placement. However, we do not have such knowledge ahead of time as the data center is shared between many jobs and free SSD capacity fluctuates over time. Thus, a solution that can apply under varying SSD capacities is needed. We now list our baselines.

\subsection{FirstFit: Static Placement.} Production systems commonly perform HDD/SSD tiering using FIFO or LRU-style heuristics \cite{yang2023fifo2, eytan2020s}. We implement a representative instance of this approach. We try to place jobs on SSD in the order of their start times, checking jobs' peak space usage and only placing jobs on SSD that fit in the available SSD capacity. This optimizes \textsc{TCIO} when unlimited SSD is available but can significantly increase TCO when SSD capacity is limited or expensive.

\subsection{Heuristic: Practical Adaptive Placement.} Recently, heuristics that can dynamically adapt to workloads have been introduced, striking a balance between dynamically learning workload behavior and avoiding the practicality issues in \Cref{ssec:production}. We emulate the state-of-the-art placement approach from \cite{yang2022cachesack}. It focuses on a slightly different problem (SSD read cache admission), but can be adapted for our placement task. We use this approach as a stand in for the closest practical approach to a learning-based baseline.

The approach starts from a set of categories associated with storage requests and then constructs a per-category admission policy based on dynamic behavior. In our experiments, we use the job's \emph{ID} as the category. For each job category, the approach measures space usage and TCO savings. We rank the categories by their TCO savings and add categories into an \emph{admission set} until the selected category's historical space usage reaches the SSD capacity. When a new job arrives, we decide to place it on SSD if it belongs to the admission set. Otherwise, the job is placed on HDD.

\subsection{ML Baseline: Lifetime Prediction-Based.} We include another closely related ML approach \cite{zhou2021learning}, which models storage problems in data centers as distribution prediction problems. We follow the paper's SSD/HDD tiering case study to predict the mean ($\mu$) and standard deviation ($\sigma$) of a file lifetime. Files with a predicted lifetime ($\mu + \sigma$) shorter than the specified time-to-live (TTL) are admitted to SSD. To mitigate mispredictions, we evict any file residing in the SSD for longer than $\mu + \sigma$, as in the paper.

\vspace{-0.4mm}
\section{Hybrid Learning Approach}
\label{sec:sys}

A common approach to ML-driven systems is to train a model that learns to make decisions, such as whether to place a file on SSD or HDD – e.g., 
via imitation learning \citep{liu2020imitation}. However, data centers are highly dynamic environments and the optimal decision depends on external factors such as the available amount of SSD at a given point in time. Thus, a model would need to jointly learn the external environment \emph{and} the workload, which is challenging and not deployable, requires a possibly prohibitive amount of training data, and may be brittle when facing new scenarios.

To address this and our other challenges (\Cref{ssec:production}), we allow each workload to ``bring their own model''. We propose a \emph{cross-layer} learning approach that uses the model only to predict a \emph{proxy} for workload-specific characteristics and then co-designs a storage-level heuristic that turns the predictions into decisions for the current environment. This design ensures that each workload’s I/O patterns is treated as independent, preventing one workload from impacting the I/O costs of others. Specifically, we want to design a proxy that correlates with a job's TCO savings. The proxy allows predicted results to directly represent how a job's placement contributes to end-to-end cost savings.

We call this proxy metric ``importance'' and train a categorical pointwise ranking model (\emph{category model}) to learn the job's \emph{importance ranking} (\Cref{fig:overview} right). Each category represents a particular level of importance. A higher ranking indicates a more important job -- placing it on SSD saves more cost. When making a placement decision, we query the model for an \emph{importance ranking category} for a new job. We then run an adaptive category selection algorithm with dynamic feedback from the storage layer to decide which importance ranking categories to admit onto SSD.

\vspace{2.7mm}

\subsection{Features} \label{sec:features}

We train our model on application-level features from production traces. The features span execution metadata, job timestamps, allocated resources, and historical system metrics, which reflect how jobs are processed in our setting.

Each shuffle job has three main steps: data writing, sorting, and data retrieval. Workers first write raw intermediate files, which are then organized into sorted intermediate files by sorters. Finally, workers retrieve the required data from the sorted files back into memory. These steps can overlap in time, depending on the job setup. To represent job I/O density, we utilize internal job-related information from the framework, which breaks down data into buckets that are assigned to workers for execution. Buckets help distribute the workload evenly and improve parallel writing efficiency.

The features we choose (\Cref{app:tab:detail-feature} in \Cref{app:model-features}) capture how these steps are executed (allocated resources, execution metadata). Execution metadata is formatted as strings that detail execution-related names, paths and targets. Key elements are separated by non-alphanumeric characters. We treat execution metadata as a sequence of substrings representing the key elements (\Cref{app:tab:example-feature} in \Cref{app:model-features}). Since executions may run periodically, we also include the weekday and hour of the day of a job's start time. Allocated resource information represents resources assigned to the job by the cluster scheduler, before it starts execution. However, specific details regarding resource distribution, such as the assignment to SSD or HDD, are not determined at this stage. In addition, we also incorporate properties of previously completed jobs from the same user's pipelines, including the past $\textsc{TCIO}$, job lifetime, and size. For a detailed description of these features, please refer to \Cref{app:model-features}.

\subsection{Model Design}\label{sec:model-feature}

\begin{figure}[t]
    \centering
    \includegraphics[width=\linewidth]{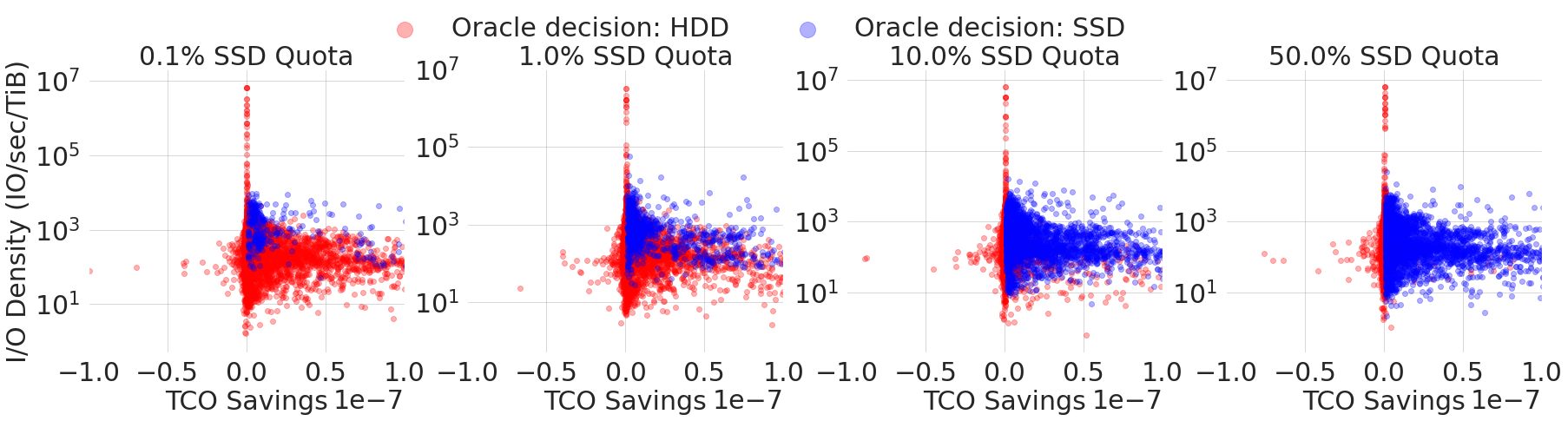}
    \caption{I/O density\xspace and TCO savings of each job (color shows oracle placement decision when optimizing for TCO). Tested under different SSD quota.}
    \label{fig:distribution-tco-iodensity}
\end{figure}

We use gradient boosted trees instead of neural networks (which are much more expensive and less interpretable) or lookup tables (which sacrifice accuracy). We build on the Yggdrasil Decision Forests framework (YDF) \cite{GBBSP23}. Our model is trained on the features in \Cref{sec:features}.

\textbf{Quantifying Job Importance.} Our goal is for the model to determine each job's \emph{importance}, which is equivalent to the expected cost savings. We first design a way to represent this \emph{importance}. We examine the oracle placement (from \Cref{sec:prob}) under different SSD capacities. We expect the most important jobs to be admitted by the oracle even under extremely limited SSD capacity, and as the SSD capacity increases, less important jobs are admitted.

For each job, we compute binary placement decisions (SSD/ HDD) from the oracle with different SSD capacity limits. In \Cref{fig:distribution-tco-iodensity}, we show how oracle decisions correlate with TCO savings and \emph{I/O density\xspace}, which denotes the total I/O across the job lifetime divided by its maximum storage footprint. As the SSD capacity increases, more jobs wıth lower I/O density are chosen for SSD. Since the oracle optimizes for TCO, jobs with negative TCO savings if put on SSD should never be selected. Further, if two jobs have the same I/O, small and short-lived jobs are preferred as they occupy less SSD capacity for less time. This suggests that predicting the \emph{sign of TCO savings} and \emph{I/O density} is a good proxy for job importance: negative TCO saving jobs are least important; jobs with higher I/O density are more important.

\textbf{Label Design.} Predicting precise values of these properties has been shown to be challenging, even in theoretical works – e.g., lifetime in \cite{zhou2021learning}. Rather than treating the importance prediction problem as regression, we choose a categorical pointwise ranking model, which groups jobs with similar (TCO savings, I/O density) into the same category (i.e., importance ranking class). The idea of framing the output prediction issue as a classification task is commonly adopted in the fields of image and audio analysis \cite{kang2020contragan, lee2009unsupervised}.

To pick the specific categories, we need to take into account that our goal is for these categories to provide a ranking of the ``importance'' of placing each job on SSD. First, negative TCO saving jobs should have the lowest ranking and we set aside one category specifically for these jobs. For the remaining categories, our goal is to cluster jobs by their I/O density. We found that both linear and logarithmically spaced categories would result in a heavily imbalanced data set (\Cref{fig:distribution-tco-iodensity}). Therefore, we choose the categories so that they evenly divide the training set by I/O density (e.g., top 10\%, top 20\%, top 30\%, etc.). For a model with $N$ categories, this results in the following category labels, given TCO savings $x.m$, I/O density $x.n$, and training set size $D$:
\begin{equation*}
\textnormal{C}(x) =
\begin{cases}
0, & \text{if } x.m < 0 \\
k, & \text{if } x.n \in ( \text{top } \frac{N-k}{N-1}*D, \text{top } \frac{N-k-1}{N-1}*D ] \\
 & \quad \text{and } x.m \ge 0
\end{cases}
\end{equation*}
\subsection{Adaptive Category Selection Algorithm}
\label{sec:adaptive-algo}

We now discuss how our cross-layer design combines the learned category model and a heuristic-based algorithm for online data placement. As mentioned in \Cref{sec:sys}, the model's category prediction is independent of the SSD capacity. How to select the categories to place on SSD under varying SSD capacities is non-trivial. A simple approach is to fix the admittable categories and always only place jobs predicted in these categories onto SSD. However, as shown in \Cref{fig:distribution-tco-iodensity}, when we have larger SSD capacities, we want to dynamically admit more jobs compared to the smaller capacities case -- that is, more categories.
\begin{algorithm}[t]
\caption{Adaptive Category Selection Algorithm}
\label{alg:adaptive-algo}
\begin{algorithmic}[1] \small
    \INPUT model $M_N$, $X=\langle x_0, \dots, x_n \rangle$, $t_w$, $T_\textsc{SpilloverTCIO}$, $t_l$.
    \STATE Initialize $t_d = 0$, $\textsc{ACT} = 1$ and $X_h = \emptyset$.
    \FOR{$x_i$ in $X$}
        \STATE Get the current time stamp as $x$'s arrival time $t_i = x_i.t_a$
        
        \algorithmicif{ Last admission decision is expired: $t_i \ge t_d + t_l$} 
        \algorithmicthen
            \STATE \quad Update look back window endpoint $w_s, w_e = t_i - t_w, t_i$
            \STATE \quad Remove expired jobs:

            $X_h = X_h - \{x_j | x_j.t_a <= w_s\}$ 
            \STATE \quad Update the spillover percentage from $X_h$: \\ \quad $h_\textsc{SpilloverTCIO} = P_\textsc{SpilloverTCIO}(X_h, t_i)$ 
            
            \quad \algorithmicif{ $h_\textsc{SpilloverTCIO} < T_l$}
            \algorithmicthen
                 \STATE \qquad $\textsc{ACT} = \max(N-1, \textsc{ACT} + 1)$
             \algorithmicend
            
            \quad \algorithmicif{ $h_\textsc{SpilloverTCIO} > T_u$}
            \algorithmicthen
                \STATE \qquad $\textsc{ACT} = \min(1, \textsc{ACT} - 1)$
            \algorithmicend
            \STATE \quad Update decision making time $t_d = t_i$
        \algorithmicend
        \STATE Infer the predicted category $C_i = M_N(x_i.\text{features})$
        
        \algorithmicif{ $C_i \ge \textsc{ACT}$}
        \algorithmicthen{ \STATE \quad Place $x_i$ onto SSD}
        \algorithmicelse{ \STATE \quad Place $x_i$ onto HDD}
        \algorithmicend
        \STATE Add the job into the observation history $X_h = X_h \cup x_i$
    \ENDFOR
\end{algorithmic}
\end{algorithm}
\begin{table}
\scalebox{0.85}{
\begin{tabular}{@{}c|c@{}} 
Symbol/Notation & Meaning \\
\hline
$X=\langle x_0, x_1, \dots, x_n \rangle$  & job sequence \\ 
$x.\text{features}$ & job features available before execution \\
$x.\textsc{Dev}$ & job scheduled device ($0/1$ for HDD/SSD) \\
$t_a, x.t_a$ & job arrival time \\
$t_s, x.t_s$ & job spillover time \\
$t_e, x.t_e$ & job end time \\
$\textsc{TCIO}_\text{HDD}(t)$ & job \textsc{TCIO}\xspace if put onto HDD until $t$ \\
$\textsc{SpilloverTCIO}(x, t)$ & job spillover \textsc{TCIO}\xspace at $t$\\
$P_\textsc{SpilloverTCIO}(X, t)$ & jobs spillover \textsc{TCIO}\xspace percent at $t$\\
\hline
$t_w$ & look back window time length \\
$t_l$ & admission decision effective time length \\
$T_\textsc{SpilloverTCIO} = [ T_l, T_u ]$ & spillover tolerance range \\
\hline
$t_d$ & the last placement decision making time \\
$X_h$ & job observation history \\
$\textsc{ACT}$ & admission category threshold $(\le N - 1)$ \\ 
\hline
$M_N$ & decision tree model with $N$ categories \\
\end{tabular}
}
\caption{Algorithm notation.}
\label{tab:notation}
\end{table}

\textbf{Algorithm Overview.} Our algorithm makes its admission decisions based on real-time feedback regarding SSD utilization. When observing limited SSD capacity, we gradually decrease the number of categories to admit. Otherwise, we admit more categories. Since our category model predicts the ``importance ranking'' of jobs, admitting fewer categories naturally leads to admitting the most important jobs. Admitting more categories means that we broaden the admission set by adding less important but still cost saving jobs. We use a sliding category admission threshold to determine which predicted categories get placed on SSD.

\textbf{SSD Usage Approximation.} In cloud data centers, the actual SSD capacity varies among clusters of machines, which is also hard to directly approximate in practice. The criteria for determining whether an SSD is nearly full (i.e. cannot fit more jobs) or not are influenced by workload patterns. Thus, we introduce a metric to unify the measurement of SSD capacity usage across clusters and workloads through job behavior observation. Given a sequence of jobs, $X=\langle x_0, x_1, \dots, x_n \rangle$, we define the \textit{spillover $\textsc{TCIO}$ percentage}, $P_{\textsc{SpilloverTCIO}}{(X, t)}$, to measure the portion of all job $x_i$'s $\textsc{TCIO}$ that is scheduled to be put onto SSD but ends up on HDD due to the fact that the SSD has already reached its full capacity at timestamp $t$ (notation in \Cref{tab:notation}):
\begin{equation*} \small
P_{\textsc{SpilloverTCIO}}(X, t) = \frac{\sum_{x_i \in X} \textsc{SpilloverTCIO}(x_i, t)}{\sum_{x_i \in X} x_{i}.\textsc{Dev} \cdot x_{i}.\textsc{TCIO}_\text{HDD}(t)}
\end{equation*}
where $\textsc{SpilloverTCIO}(x, t)$ is the amount of spill over $\textsc{TCIO}$ of a job $x$ at time $t$, if spillover starts at $t_s$:
\begin{equation*}  \small
\begin{cases}
\frac{t - t_s}{t - t_a}\textsc{TCIO}_\text{HDD}(t), &\text{if } t_s \text{ exists and } t_a \le t_s \le t \\
0.0, &\text{Otherwise}.
\end{cases}
\end{equation*}
Intuitively, $\textsc{SpilloverTCIO}$ measures the amount of the job's intended \textsc{TCIO} savings that are not realized. A large $P_\textsc{SpilloverTCIO}$ means that many jobs fail to be scheduled onto SSD, which indicates that the SSDs are nearly full.

\textbf{Algorithm Design.} We now introduce the adaptive category selection algorithm in \Cref{alg:adaptive-algo} with notation available in \Cref{tab:notation}. In the algorithm, we keep track of an observation history $X_h$, which contains all the jobs starting within a fixed look back window, and calculate the $\textsc{SpilloverTCIO}$ within the history, $h_{\textsc{SpilloverTCIO}}$. Then, we adaptively adjust the \emph{admission category threshold} (ACT) based on the observed $P_\textsc{SpilloverTCIO}$ -- if $P_\textsc{SpilloverTCIO}$ is larger than ACT, we increase the threshold to admit fewer categories. One issue of a dynamic control system of this kind is that ACT may change drastically. We combine two mechanisms to smooth the ACT change:
\begin{itemize}[itemsep=1pt,wide = 0pt,topsep=0.5pt]
    \item We use a spillover tolerance range, $T_\textsc{SpilloverTCIO}$, within which the ACT remains unchanged. Avoiding large adjustments, if $P_\textsc{SpilloverTCIO}$ falls below the range lower bound, we decrease the threshold by 1; if $P_\textsc{SpilloverTCIO}$ exceeds the upper bound, we increase the ACT by 1.
    \item This ACT update is triggered only at job arrivals and at fixed decision intervals of $t_l$ seconds, rather than on every job arrival. This reduces the frequency of threshold changes.
\end{itemize}

When we designed the algorithm, we discovered that considering all the jobs \emph{starting} within the look back window can result in a more accurate estimate of the latest SSD usage information than using all the jobs \emph{overlapping} the look back window. We think this could be the result of jobs with a long lifetime having an outsize effect in such a setting.
\section{Evaluation}
\label{sec:eval}

We study the following research questions for evaluation:
\begin{enumerate}[label=\upshape\bfseries RQ\arabic*:, wide = 0pt, itemsep=0pt,topsep=0pt] 
    \item What is our method's performance when integrated into \google's system?
    \item What are our method's TCO and \textsc{TCIO} savings?
    \item What are our method's TCO savings under different SSD space constraints?
    \item How does our method generalize across workloads, or perform with new users and pipelines?
    \item Is our ML model practical? Which features contribute most to learning? 
\end{enumerate}

As described in \Cref{sec:intro}, most of our results are based on a diverse set of workloads written against a shared data processing framework. To show generality, we also demonstrate that the approach works with other workloads. These results, as well as additional experiments and sensitivity analyses, are shown in \Cref{app:results}.

\subsection{Experimental Setup}

\textbf{Metrics.} We evaluate our performance using two metrics: \textit{TCO savings percentage} and \textit{\textsc{TCIO} savings percentage}. As described in \Cref{sec:prob}, TCO includes the total expenditure of maintaining a storage systems (such as I/O cost, SSD wearout, etc). TCO savings are relative to the total TCO if all jobs are put on HDD. \textsc{TCIO} measures the actual I/O cost without calculating the SSD wearout and network cost. Given that the SSD wearout cost could differ in different contexts, we believe showing \textsc{TCIO} can help understand the savings purely from an I/O perspective. Similar to TCO savings, we show \textsc{TCIO} savings as the percentage of \textsc{TCIO} savings over the \textsc{TCIO} if all jobs are put on HDD.

\textbf{Data Collection \& Model Training.} We collect  production traces from $\google$ that consist of the historical execution log of the data processing framework and the I/O traces from the distributed storage system. These traces contain jobs' metadata and post-execution measurements, such as \textsc{TCIO}. \Cref{sec:model-feature} explains the features we use. Our training and test dataset each contains one week's data, which are collected from a contiguous two-week time span. We include both read-heavy and write-heavy jobs, reflecting the diversity of workloads across the hyperscale fleet. Block sizes of these workloads vary from several KiB to MiBs.

A key trade-off in our approach is the granularity of model training. We could train one model for every binary, or a joint model that can be used across multiple workloads. To scale to the large number of workloads in our data set, we use the latter approach. Since similar workloads are often run within the same server cluster/data center, we use clusters as the granularity of model training, jointly training one model per cluster and using it for all workloads in this cluster. This has sufficient accuracy (\Cref{sec:model-analysis}), but nothing precludes us from using a finer or coarser granularity. We use a 15-class gradient-boosted trees model with 300 trees at maximum and a max depth of 6 for all of our models.

\textbf{End-to-end System Integration.} We develop a prototype and deploy it in our production data processing framework and distributed storage system at \google. In the prototype, we first follow \Cref{sec:features} to train the per-cluster model offline. Our trained models are located inside the data processing framework during execution. There is no need to modify user application code. On the compute nodes and inside each job process, the computation framework collects required features and loads the model to perform inference, to generate a categorization result before opening files for writing. The categorization results are passed to the storage cache server, which makes real-time decisions for placement on HDD or SSD. Metrics are collected during the process to evaluate TCO and \textsc{TCIO}.

\pagebreak[4]

\textbf{Large-Scale Simulation Setup.} We conduct extensive simulation using real production traces at the scale of a cluster. The simulations allow us to perform detailed study of performance and trade-offs at a large scale. Our simulation executes job placement on either SSD or HDD. If a job is placed on SSD but only partially fits, the remaining portion of the job spills over to HDD after filling the available SSD capacity. For experiments varying SSD capacities, we initially set the SSD constraint to infinity to determine the cluster's maximum space usage. We then simulate different scenarios by varying the SSD space quota.

\textbf{Methods Compared.} We compare 7 methods (\Cref{sec:prob}): \emph{FirstFit} (simple heuristic), \emph{Heuristic} (advanced heuristic, a modified \cite{yang2022cachesack}), \emph{ML Baseline} (following \cite{zhou2021learning}),  \emph{Adaptive Hash} (our method without ML models), \emph{Adaptive Ranking} (our method), \emph{Oracle \textsc{TCIO}} (best theoretical bound when optimizing \textsc{TCIO}), and \emph{Oracle TCO} (best theoretical bound when optimizing TCO).

\begin{figure}[t]
   \centering
      \includegraphics[width=\linewidth]{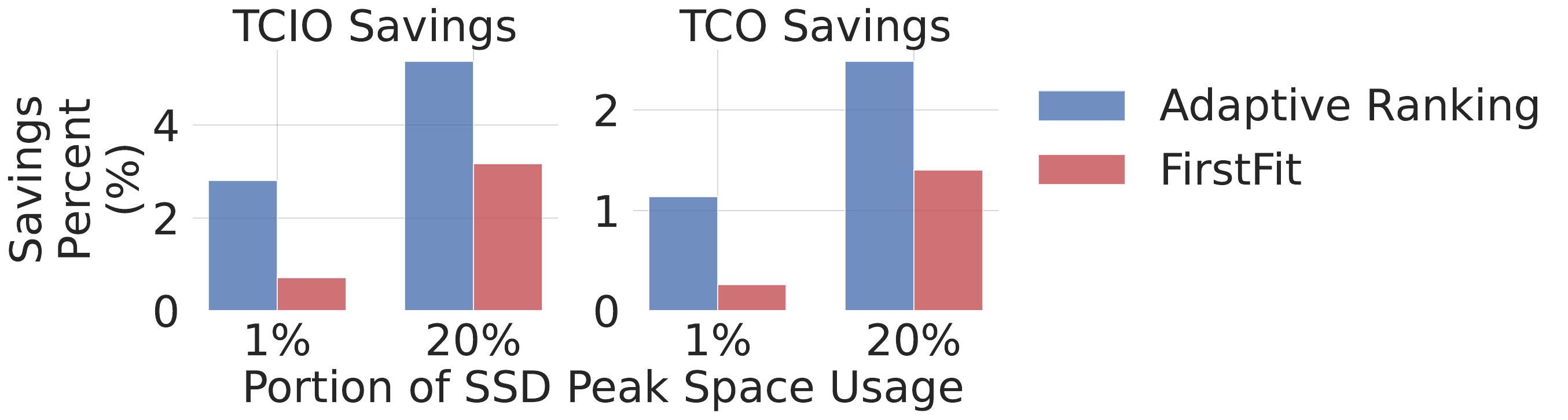} 
      \caption{Prototype results.} \label{fig:prototype-results}
\end{figure}

\begin{figure}[t]
    \centering
    \includegraphics[width=\linewidth]{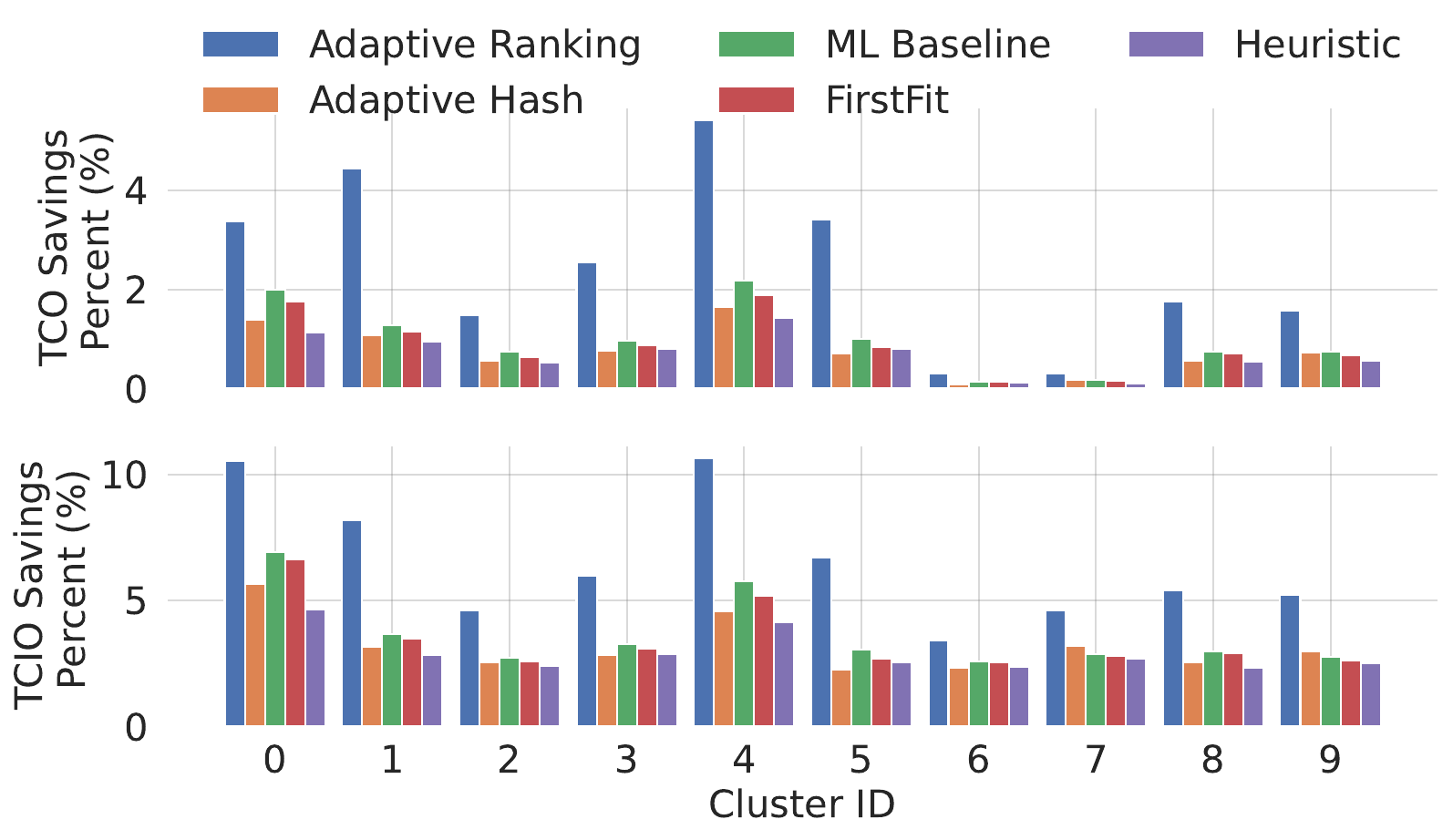}
    \caption{TCO savings (top) and \textsc{TCIO} savings (bottom) from different clusters with fixed SSD quota.}
    \label{fig:per-cluster-savings}
\end{figure}

\begin{figure*}[t]
   \begin{minipage}[b]{0.5\textwidth}
     \centering
     \includegraphics[width=0.76\linewidth]{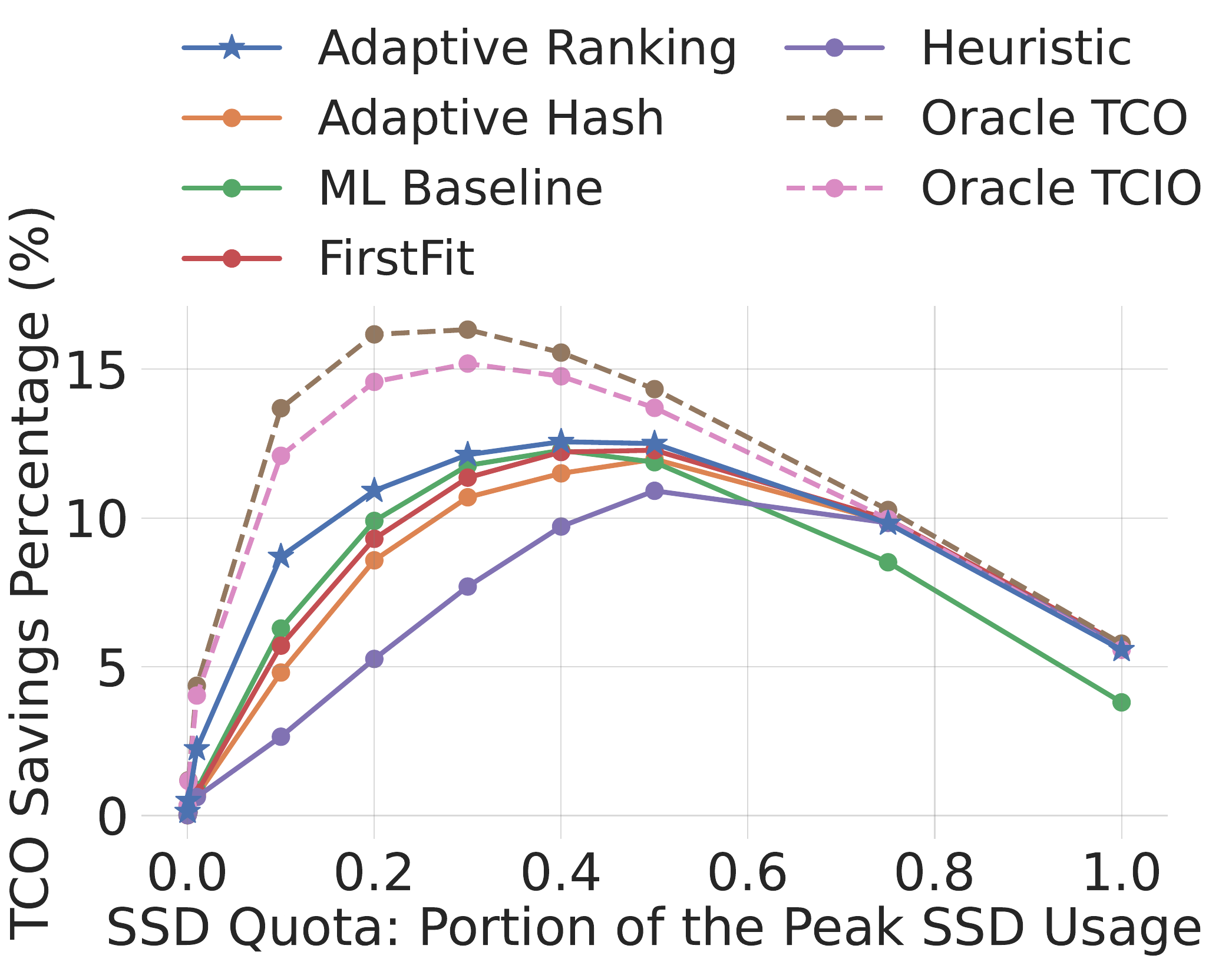}
     \caption{TCO savings.}
     \label{fig:tco-curve}
   \end{minipage}
   \begin{minipage}[b]{0.5\linewidth}
     \centering
     \includegraphics[width=0.78\linewidth]{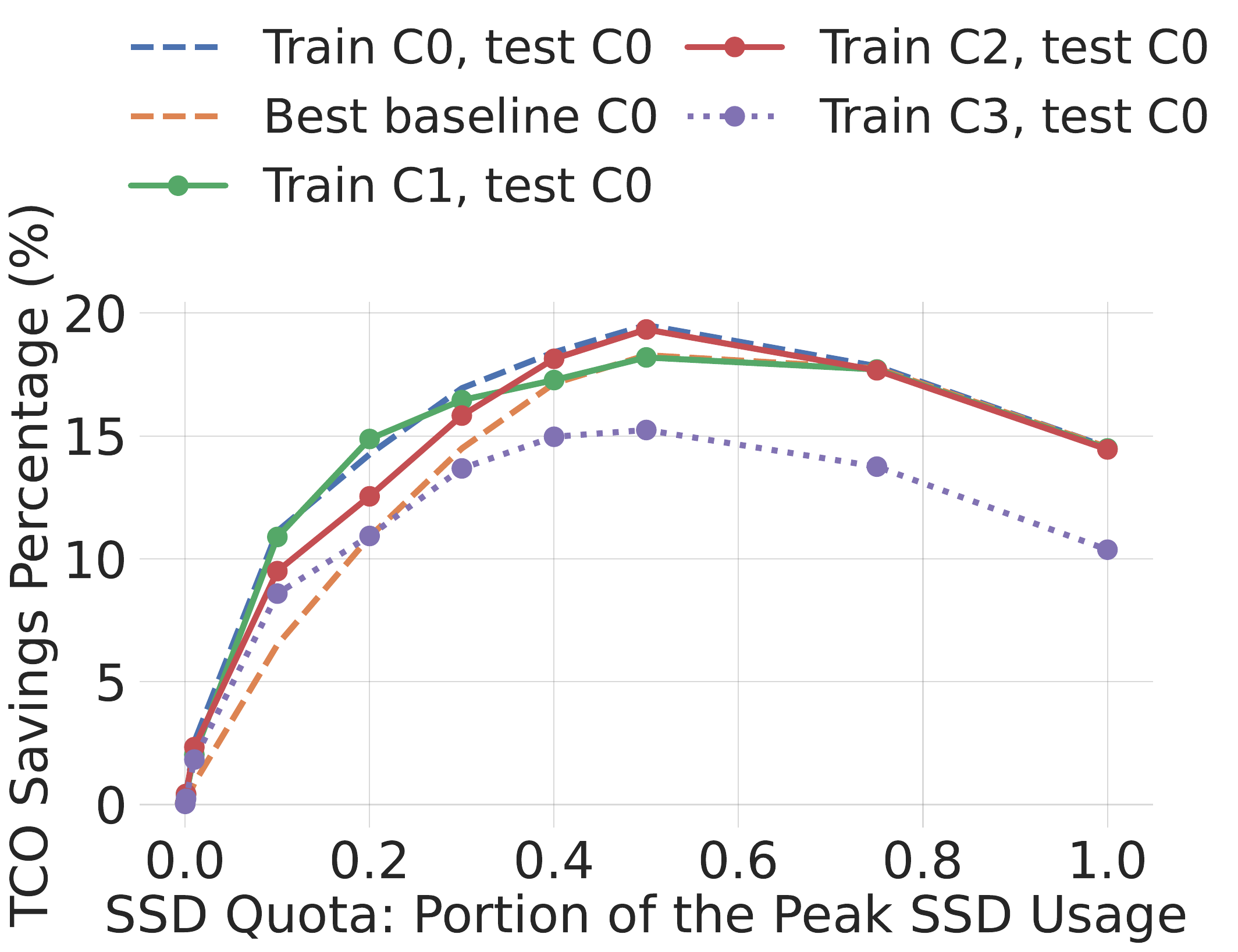}
     \caption{Workload generalization.}
     \label{fig:workload-gen}
   \end{minipage}
\end{figure*}
     
\begin{figure}[t]
  \begin{minipage}[b]{\linewidth}
  \centering
    \begin{subfigure}{0.49\linewidth}
        \centering
        \includegraphics[width=0.9\linewidth]{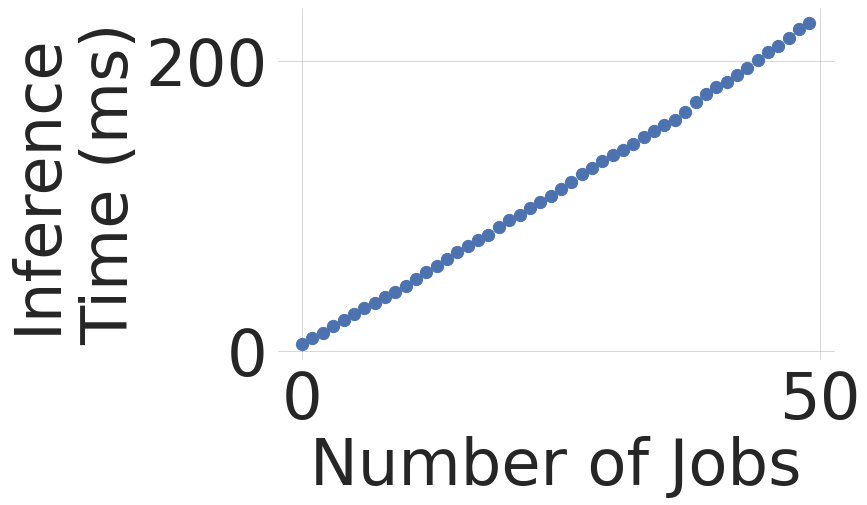}
        \caption{Running time}
        \label{fig:model-running-time}
    \end{subfigure}
    \begin{subfigure}{0.49\linewidth}
        \centering
        \includegraphics[width=0.9\linewidth]{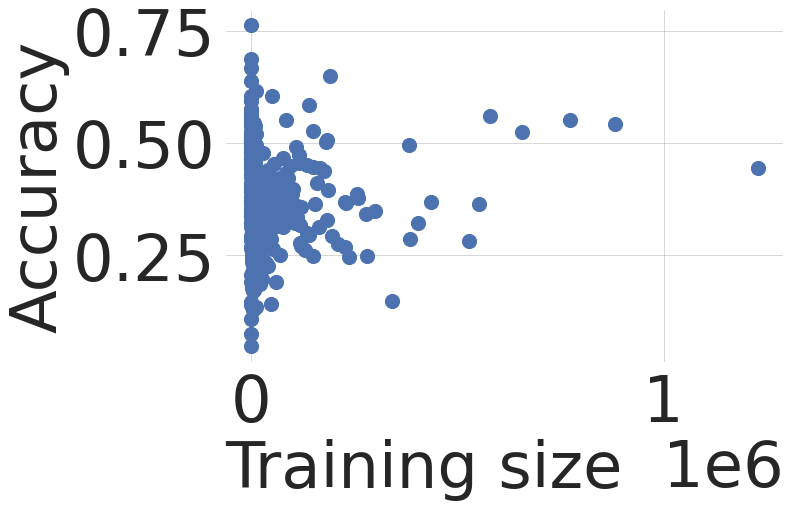}
        \caption{Accuracy}
        \label{fig:accuracy-size}
    \end{subfigure}
    
    \begin{subfigure}{\linewidth}
        \centering
        \includegraphics[width=0.85\linewidth]{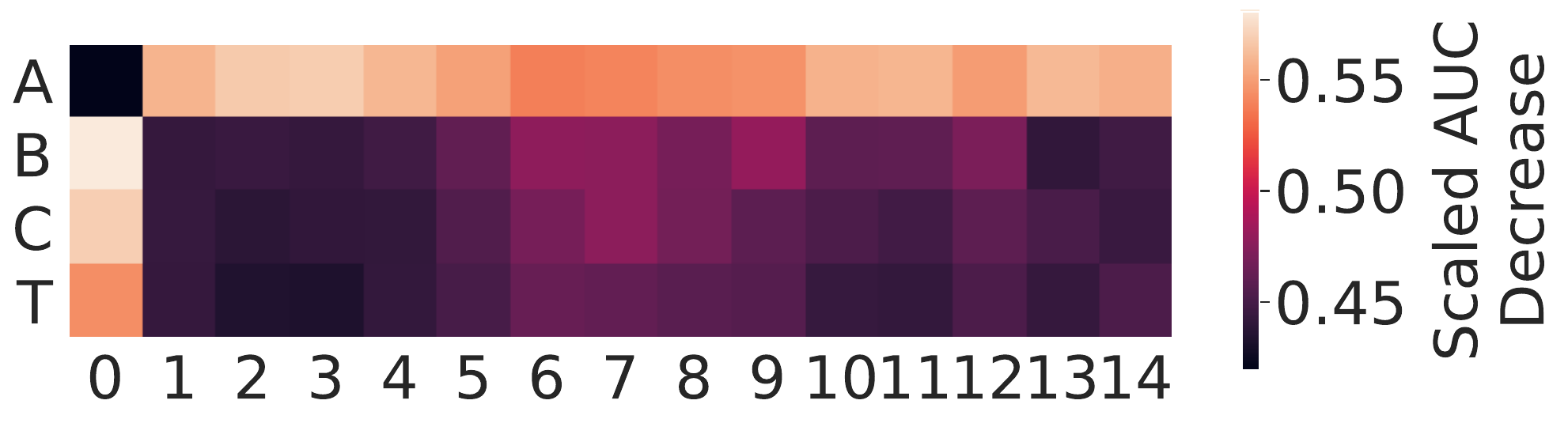}
        \caption{Importance of different feature groups}
    \label{fig:feature-importance}
    \end{subfigure}
    \caption{Model analysis.}
  \end{minipage}
\end{figure}

\subsection{Integration in Real Systems [RQ1]}
\label{subsec:eval-integration}

We pick one cluster and implemented two methods in \google's storage system: FirstFit and Adaptive Ranking (ours). A range of data pipelines are selected to generate I/O workloads in this prototype. These pipelines cover a variety of large dataset processing tasks, which generate a wide range of I/O workloads with different intensity and throughput. One category of pipelines is more cost-effective when using HDD, while the other category is more cost-effective to run on SSD. These pipelines are executed continuously in a production cluster, akin to real production pipelines. We set up a dedicated SSD cache so that more precise and disturbance-free results can be measured. A total of 320 worker servers are used to execute the workloads, which includes 16 pipelines and 1024 shuffle jobs in total. The pipelines' combined peak storage usage is 3.6\,TiB.

We set up two scenarios where the SSD quota is 1.0\% and 20\% of the peak theoretical SSD usage limit (\Cref{fig:prototype-results}) respectively, which are common in real-world deployments. For the 1.0\% case, our algorithm shows 1.14\% TCO savings (\textbf{4.38$\times$} over FirstFit). Our method gives 2.48\% TCO savings (\textbf{1.77$\times$} over FirstFit) in the 20\% of the peak workload space usage. The \textsc{TCIO} savings indicate a similar pattern: Adaptive Ranking is 3.90$\times$ and 1.69$\times$ over FirstFit in the two SSD quota cases respectively. 

The end-to-end prototype demonstrates the viability of our design. The measured savings of Adaptive Ranking and improvements over the baseline are in line with the performance in the large-scale simulation studies in \Cref{subsec:eval-overall-savings} and thus validate our simulation methodology.

\subsection{Overall Savings [RQ2, RQ3]}
\label{subsec:eval-overall-savings}

We pick 10 clusters with large TCO to evaluate overall savings. Each of these clusters has thousands of machines. There is considerable variation between workloads within clusters, including video processing, ML, and database queries. Furthermore, the distribution of applications is uneven among clusters. To show performance across different clusters, we fix the SSD quota at $1.0\%$ of the peak SSD space usage and show savings per workload (\Cref{fig:per-cluster-savings}). Our method (Adaptive Ranking) can save over 3.47$\times$ at maximum (2.59$\times$ on average) compared with the best baselines in terms of TCO savings. The \textsc{TCIO} savings follow a similar pattern. Typically, the \textsc{TCIO} savings increase with SSD quota because SSD cost is not considered. In comparison, the TCO savings initially increase as SSD quota goes up but drop when SSD quota is very large due to high maintenance costs of SSDs. We consider our approach as an effective solution especially when SSD space is limited.

We also evaluate the TCO savings when SSD quota varies. In practice, we want an approach that can adapt to external factors (such as changing SSD quota). Our method consistently saves more TCO than baselines, especially in limited SSD quota cases (\Cref{fig:tco-curve}). The gap between our method (adaptive ranking) and adaptive hash (non-ML) clearly indicates the necessity of our category model. The gap between oracle (best possible in theory) and our method also indicates the remaining headroom for future work.

\subsection{Generalizability [RQ4]}

We explore generalizability of our method across users and workloads. In practice, good generalizability is necessary as infrastructure, user behaviors, workloads, etc.\ change over time. First, we evaluate the generalizability across workloads  (\Cref{fig:workload-gen}). We train one category model for each cluster $C0/C1/C2/C3$, and evaluate their performance on $C0$. $C3$ is a special cluster that only runs certain workloads that are rare in other clusters. We thus find that our method can adapt to unseen workloads, except for outliers.

\begin{figure}[t]
\begin{subfigure}{\linewidth}
    \centering
    \includegraphics[width=\linewidth]{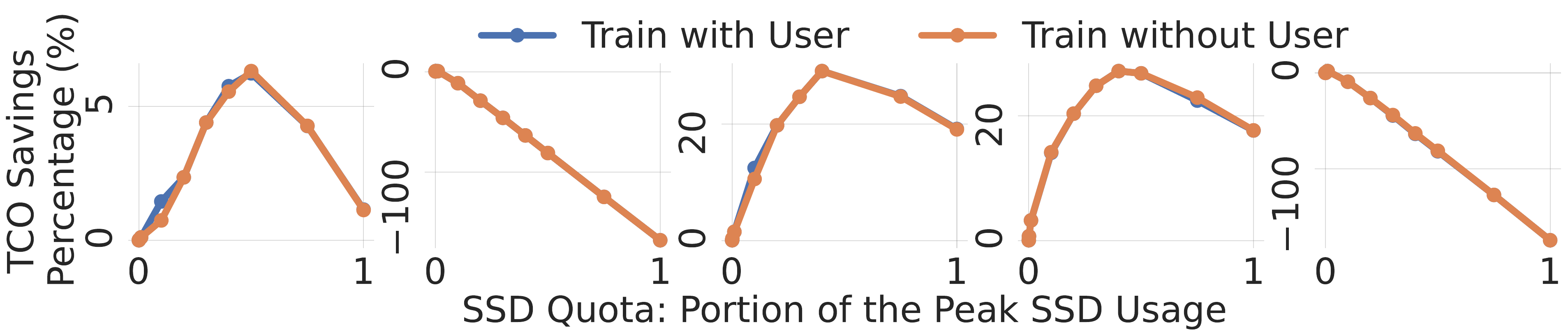}
\end{subfigure}
\begin{subfigure}{\linewidth}
    \centering
    \includegraphics[width=\linewidth]{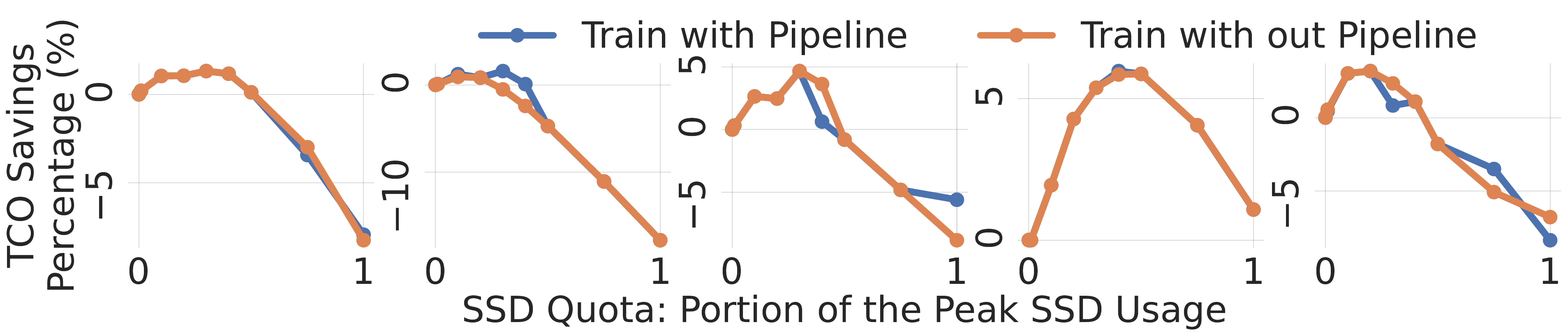}
\end{subfigure}
\caption{Generalization on new users (upper) and new pipelines (lower). Each figure is for one cluster.}
\label{fig:user-pipeline-generalization}
\end{figure}

Second, we evaluate the performance on new incoming users' jobs or jobs from unseen pipelines. We pick the second-largest TCO consuming user and pipeline, which are in different clusters. In evaluation, we compare two training methods: 1) We train the category model with the historical workloads including the user or the pipeline. 2) We train the category model with the historical workloads excluding the user or the pipeline. We evaluate the TCO savings curve under five clusters and show that our method can achieve similar TCO savings on new users or pipelines as in the case where the users or pipelines are in the training set (\Cref{fig:user-pipeline-generalization}). The blue line (user or pipeline included in the training) and the orange line (user or pipeline excluded in the training) achieve similar savings online.

\subsection{Model Analysis [RQ5]}
\label{sec:model-analysis}

Another practical requirement for ML models is their execution latency and explainability. We show the accumulated inference time of 50 jobs in \Cref{fig:model-running-time}, where inference takes about 4 ms per job (considerably lower compared to the 99ms of the Transformer model in \cite{zhou2021learning}), fast enough for making online placement decisions. Note that our ML model invocation is currently an unoptimized prototype implemented in Python. Potential efficiency improvements could be achieved through further performance optimization, such as using YDF's C++ binding.

We show our model accuracy in \Cref{fig:accuracy-size}. We compare the relationship between model accuracy and the training size across all workloads. The average top-1 accuracy is $0.36$ for 15-category classification model and we do not spot a strong correlation between training size and accuracy, which indicates that large data size may not be strictly required.

Next, we analyze how the model makes decisions by assessing feature importance across feature groups, as depicted in \Cref{fig:feature-importance}. Feature group significance is represented by color intensity, with lighter shades indicating higher importance. As discussed in \Cref{sec:model-feature}, the features are divided into four groups: A (Historical system metrics), B (Execution metadata), C (Allocated resources), and T (Job timestamp). 

To quantify the importance of features in predicting each category, we perform a binary prediction analysis (whether a job belongs to the category or not) for each category. For each feature, we measure the decrease in the area under the ROC curve (AUC) when that feature is excluded from binary prediction tasks. This approach helps us understand how the absence of a feature affects the model's predictive performance. These scores are normalized for comparability within each category. We calculate and present average importance scores for each feature group. In our model, the category $0$ is for negative TCO savings and the remaining categories are associated with the ranking of I/O density. Our findings reveal that historical system metrics significantly influence the prediction of I/O density rankings. In contrast, the start time and execution metadata are more critical for predicting whether a job's TCO saving is negative.

\begin{figure}[!t]
     \centering
     \includegraphics[width=0.75\linewidth]{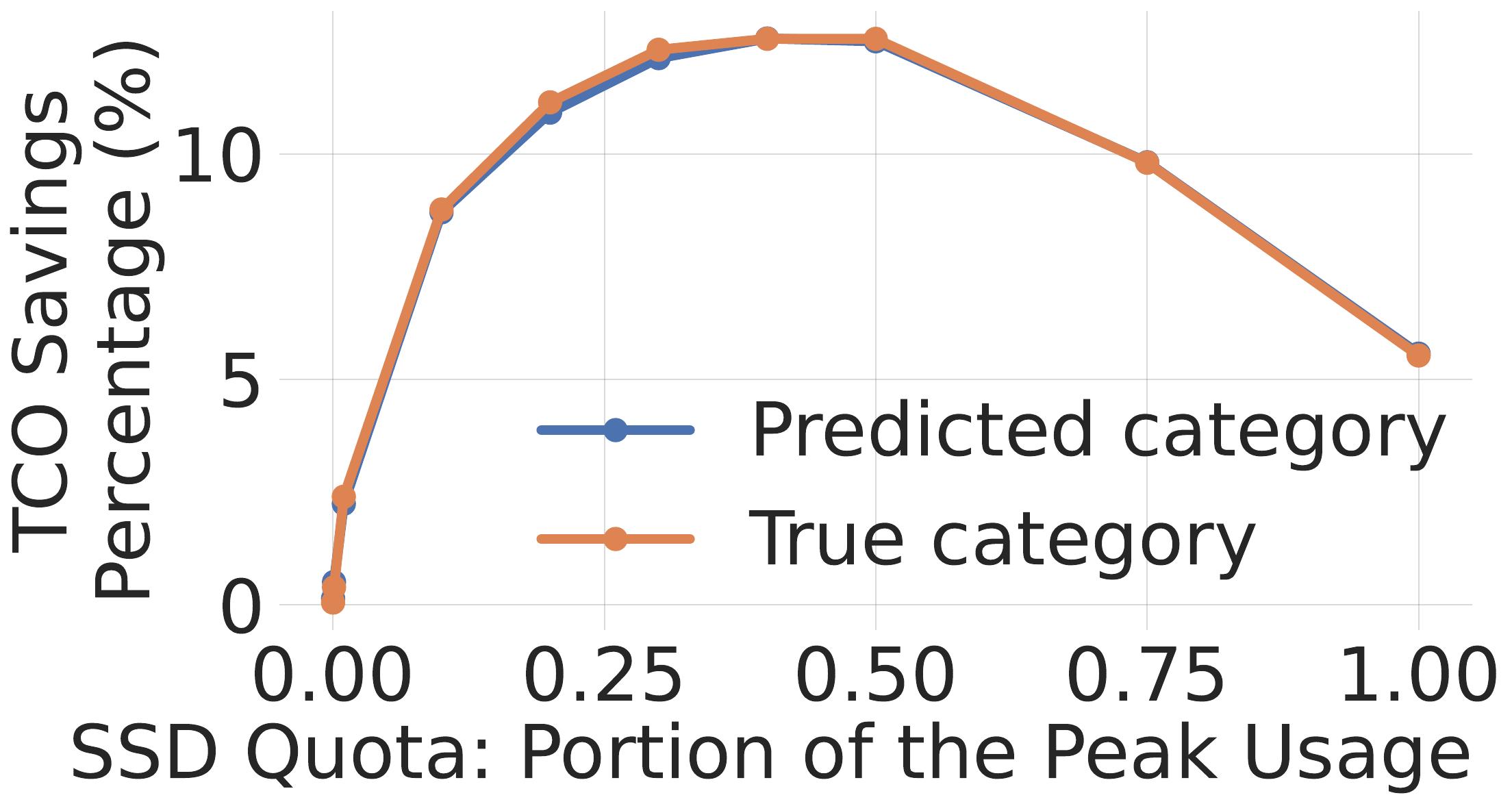}
    \caption{Comparison with using true category.}
    \label{fig:true-category}
\end{figure}

We also find that our end-to-end savings may not benefit from more accurate models in \Cref{fig:true-category}. Here, we evaluate the performance of the TCO savings assuming we have perfect models for classification prediction. The ``Predicted category'' is our approach. The ``True category'' is a method where we replace the category prediction described in \Cref{sec:model-feature} with the ground truth category, representing 100\% prediction accuracy. 

This observation highlights an important insight: For a given algorithm, there are diminishing returns from further improving model accuracy beyond a certain point. Future improvements in TCO savings are more likely to come from refining the design of the category itself—specifically, its ability to accurately capture job importance—and enhancing its integration with the underlying adaptive algorithm. The synergy between effective category design and the adaptive algorithm plays a crucial role in achieving optimal results.

This finding helps us rethink that the challenges of ML for Systems are not solely about improving model accuracy or learning algorithms. Instead, how the learning problem is formulated and how the ML model is applied within the system are equally important factors in achieving practical performance gains.

\section{Related Works}
\label{sec:related-work}

\textbf{ML in Storage Systems.} Prior works have shown the viability of ML for task property prediction in storage systems \cite{liu2020imitation, kaler2023deep, singh2022sibyl, akgun2023improving, vietri2018driving, chakraborttii2020improving}. \cite{hao2020linnos} leverages a small neural network to infer SSD performance with fine granularity and help parallel storage applications. \cite{zhou2021learning} tackles a related SSD tiering problem with using application-level information and distributed traces by taking inspiration from natural language processing. While the paper focuses on a specific learning problem of mapping textual metadata to storage-related properties, our work focuses on the practical designs and deployment of such models. Additionally, in contrast to existing work, we propose a more granular and practical approach by learning dedicated models for each workload in a cross-layer manner.

\textbf{Data Placement in Practice.} Though ML for systems has been widely explored in different application domains, the state of the art practical solutions for caching or tiering in storage systems are still mostly heuristic \cite{pakize2014comprehensive, raj2012enhancement, zaharia2009job, azureTiering, yang2022cachesack, yang2023fifo, yang2023fifo2, eytan2020s, zhao2023tectonic}. Another noteworthy work presents a solver-based solution for task scheduling in the setting where each task contains a list of preferred locations identified prior to scheduling. Their approach formulates the problem as a minimum cost maximum matching problem \cite{herodotou2021trident}. Although closely related to our work, as discussed in \Cref{sec:bg} and \Cref{sec:prob}, the method is not directly feasible in our context, since this approach requires clairvoyant knowledge of jobs' costs and scheduling times.

\Cref{app:related-works} has more discussion of related work.

\section{Conclusion}
\label{sec:conclude}

We present a practical approach to data placement in data centers. We identify and solve practical challenges with a cross-layer data placement solution combining application-level ML models with storage-level heuristics. Our approach shows significant TCO savings over the SOTA. We believe that our cross-layer approach presents a methodology for practical ML usage in systems beyond the data placement problem (discussed more in \Cref{app:discussion}).

{
\selectfont\textbf{Acknowledgements:} 
We want to thank Larry Greenfield, Tzu-Wei Yang, Deniz Altınbüken and the anonymous reviewers for invaluable feedback and discussions.
}
\vfill

\bibliography{main}
\bibliographystyle{mlsys2025}

\newpage
\appendix
\onecolumn
\section{Production Setup}
\label{app:production-details}

\begin{figure}
    \centering
    \includegraphics[width=0.8\linewidth]{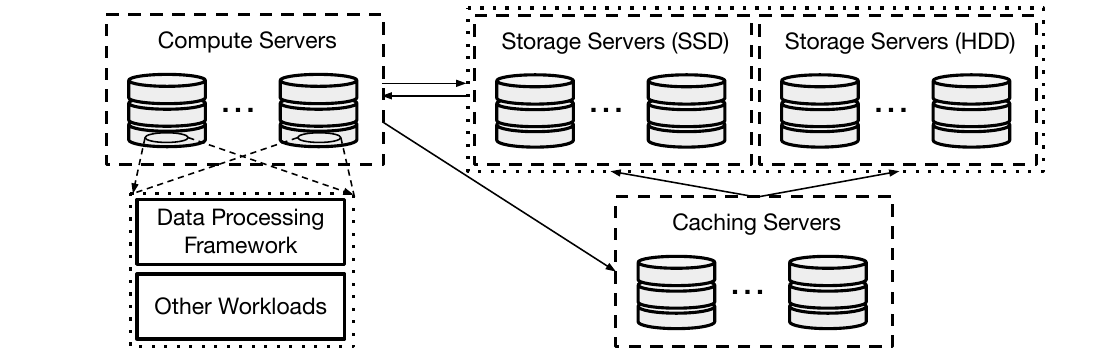}
    \caption{Production Setup Overview}
    \label{fig:prod-overview}
\end{figure}

\Cref{fig:prod-overview} provides an overview of our distributed production setup. This setup includes three dedicated sets of servers central to storage placement: 1) compute servers, which run  data processing frameworks and other workloads; 2) storage servers, which host HDD and SSD devices; and 3) caching servers, which manage SSD tiering decisions.

\section{Model Features}
\label{app:model-features}

Our design centers around the intermediate files of data processing frameworks. \Cref{sec:storage-for-data-processing-frameworks} introduces the fundamental concepts of how the data processing framework processes input data records. A distributed framework spawns workers to execute tasks. A worker is a process that runs on a server. Workers use shuffling to exchange data between them. A shuffle job is generated when the execution of the workflow reaches a step or operation that necessitates the exchange of information. As an example, \emph{GroupByKey} is a common operation across frameworks that generates one or more shuffle jobs.

At a higher level, a job comprises three steps. We assume that each worker possesses a number of data records in their working memory. In the first step, each worker writes the data they own into raw intermediate files. Accessing the data in these raw files is inconvenient because they lack a specific order. To address this issue, one or more sorters organize the data records in these files into sorted intermediate files as part of the second step. In the third step, the workers retrieve their required data from the sorted intermediate files back into their working memory, concluding the shuffle job. If feasible, these three steps can be executed concurrently, resulting in temporal overlap.

The I/O density of jobs depends on how these data records are written and read, so we are providing as much internal job-related information from the framework to the model as possible. Internally, the data a workflow needs to process is divided into buckets. A bucket is a unit of work that is assigned to a worker. Each bucket contains a set of tasks that are executed by a single worker. The number of buckets is determined by the data to be shuffled and the number of workers available. Buckets are used to ensure that work is distributed evenly across workers and that no worker is overloaded.

In the first step of a job, the worker shards the data in each bucket into shards, and each shard is assigned to a writer for being written to storage. A writer packs data into stripes and writes one stripe at a time. This enables parallel writing and faster write throughout. The feature we choose, as described in \Cref{app:tab:detail-feature}, reflects how these steps are being executed.

\begin{table*}[ht]
\centering
\scalebox{0.85}{
\begin{tabular}{l|l|l} 
Features & Feature Group & Description \\
\hline
average\_\textsc{TCIO} & Historical system metrics & Average \textsc{TCIO} of the job's historical executions. \\ 
average\_size & Historical system metrics & Average peak intermediate file size of the job's historical executions. \\ 
average\_lifetime & Historical system metrics & Average job historical lifetime. \\ 
average\_\text{I/O density} & Historical system metrics &  Average I/O density of the job's historical executions. \\
\hline
bucket\_sizing\_initial\_num\_stripes & Allocated resources & The initial number of stripes a shard is expected to be divided into.  \\
& & Each stripe contains a couple of data records. \\ 
bucket\_sizing\_num\_shards & Allocated resources & The number of shards the working set is expected to be sharded into. \\ 
bucket\_sizing\_num\_worker\_threads & Allocated resources & Number of worker threads. \\ 
bucket\_sizing\_num\_workers & Allocated resources & Number of workers in this job. \\ 
initial\_num\_buckets & Allocated resources & The initial number of buckets the job uses when it was started. \\ 
num\_buckets & Allocated resources & The number of buckets the current job actually uses. \\ 
records\_written & Allocated resources & The number of records to be shuffled for a shuffle job. \\ 
requested\_num\_shards & Allocated resources & Number of shards the current working set is requested to be sharded into. \\ 
\hline
open\_time\_dayhour & Job timestamp & The hour of the job start time. \\ 
open\_time\_seconds & Job timestamp & The second of the job start time. \\ 
open\_time\_weekday & Job timestamp & The week day of the job start date. \\ 
\hline
build\_targetname & Execution metadata & The target in the build file that is used to build the executable binary. \\ 
execution\_name & Execution metadata & A user-assigned identifier for the job. Usually set to the binary file name. \\
pipeline\_name & Execution metadata & Name of the pipeline the job belongs to. A pipeline contains multiple jobs. \\
step\_name & Execution metadata & A computer generated step identifier from the workflow's execution graph. \\
user\_name & Execution metadata & Name of the workflow step that is starting the shuffle job. \\
\end{tabular}
}
\caption{Feature details.}
\label{app:tab:detail-feature}
\end{table*}

\begin{table*}[ht]
\centering
\scalebox{0.85}{
\begin{tabular}{l|l} 
Features & Example Values \\
\hline
build\_targetname & //storage/\censor{xxxxxxx}/build\_manager:\censor{xxxxxxxxxxxxxxxxxx} \\ 
execution\_name & com.\censor{xxxxxx}.\censor{xxxxxx}.\censor{xxxxxxxxxx}.\censor{xxx}.trigger2.launcher.Main\\
pipeline\_name & \censor{xxxxxxxxxxxxxxx}\_org\_\censor{xxxxxx}\_indicator\_metrics.\censor{xxxxxxxxxxx}-dims\_prod.\censor{xxxxxxxx}.data\_importer\\
step\_name & \censor{xxx}-open-shuffle10\\
user\_name & GroupByKey-22\\
\end{tabular}
}
\caption{Feature examples.}
\label{app:tab:example-feature}
\end{table*}

\section{Additional Results}
\label{app:results}

\subsection{Evaluations with non-Data-Processing Framework Workloads}

Throughout the paper, we focus on workloads written against the same large-scale data processing framework to evaluate against a wide range of different workloads. We run additional experiments to demonstrate that our prototype and general approach is not limited to this data processing framework, but can handle any workload that supports our distributed storage system.

Note that our ``bring your own model'' approach means that workloads have a large degree of freedom in terms of producing the predictive category signals that are passed to the storage layer. We demonstrate this flexibility here, by picking diverse workloads that are entirely well-suited for SSD, or entirely well-suited for HDD (i.e., even a model that predicts the same category for each file in this workload would perform reasonably well). We use the adaptive ranking algorithm design to first develop an oracle model based on the workloads' TCO savings and I/O density, then train a model to assign file categories.

Two methods are implemented and compared for the real-world mixed workloads evaluation: FirstFit and our Adaptive Ranking. The real-world evaluation is done using a mix of workloads based on our data processing framework (referred to as ``framework workloads'') and conventional workloads (referred to as ``non-framework workloads''). Our goal is to understand how well these mixed workloads work in the real world. In our evaluation, we maintain a 1:1 framework workloads to non-framework workloads ratio in terms of generated file size footprint.

The following workloads are used for this evaluation:
\begin{enumerate}[nosep,topsep=0pt]
  \item \textbf{4 HDD-suitable framework data processing workloads}. These are data processing workloads that perform a small amount of shuffles.
  \item \textbf{4 SSD-suitable framework data processing workloads}. These are large query workloads that perform a large amount of table joints and therefore need a lot of shuffles.
  \item \textbf{10 HDD-suitable (low I/O intensity) non-framework workloads}. These are ML training  workloads with checkpointing, using the same ML framework that we used for our own models. Since these checkpoint files are kept for longer than a few hours, they are not suitable for being saved to SSD.
  \item \textbf{10 SSD-suitable (high I/O intensity) non-framework workloads}. These jobs emulate a user workflow that consists of compressing input data, generating (compressed) temporary files, uploading them to a cloud storage, and deleting the temporary files. These workloads generate hot and short lived files.
\end{enumerate}

A total of 320 worker servers are used to execute the workloads. The workloads' combined peak storage usage is 3.8\,TiB. All of the four workloads use gradient-boosted tree category models.

\subsubsection{Storage TCO and TCIO Savings}

\begin{figure}
    \centering
    \includegraphics[width=\linewidth]{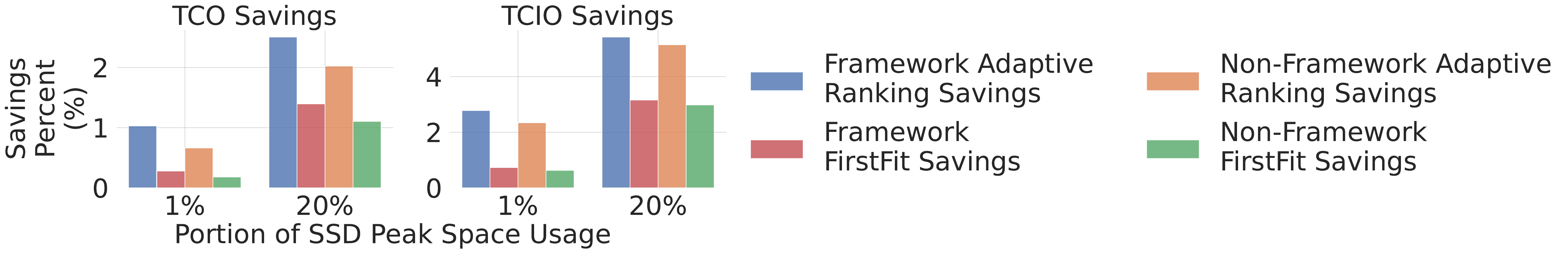}
    \caption{Prototype mixed workload savings.}
    \label{fig:prototype-mixed-workload-results}
\end{figure}

The measured TCO and TCIO of FirstFit and our Adaptive Ranking are compared with the FirstFit baseline's TCO and TCIO. The results are shown in Figure~\ref{fig:prototype-mixed-workload-results}. We see that we get significant TCO and \textsc{TCIO} savings compared to FirstFit, for both our framework and non-framework workloads. This demonstrates that our approach is not limited to workloads written in our data processing framework.

\subsubsection{Application-level Performance}

\begin{figure}
    \centering
    \includegraphics[width=.8\linewidth]{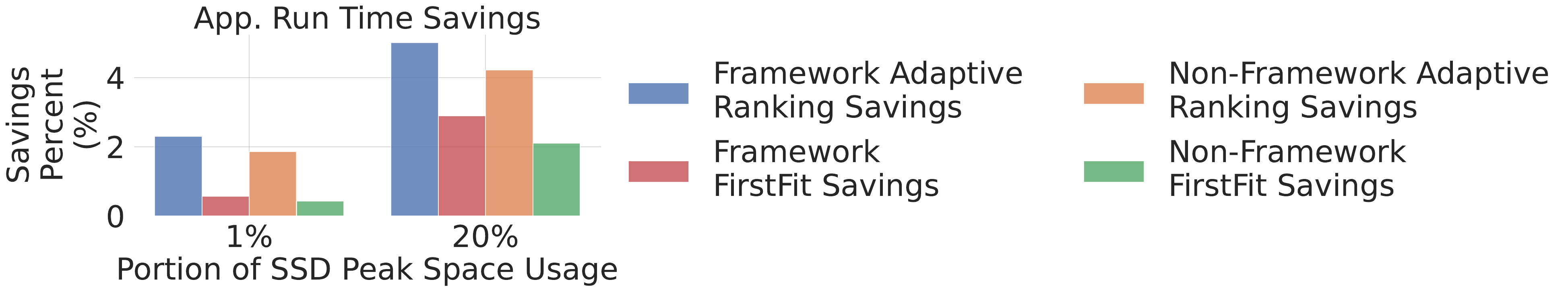}
    \caption{Prototype mixed workload application run time savings.}
    \label{fig:prototype-mixed-workload-app-level-results}
\end{figure}

We also look into the change of application-level performance introduced by our method. Since our workloads have a fixed amount of work for each execution, we measure the framework and non-framework workload overall execution time as a way to understand the application-level performance. The result is shown in Figure~\ref{fig:prototype-mixed-workload-app-level-results}. We see that the application-level performance of all workloads improves, in addition to TCO and \textsc{TCIO} savings. Most importantly, no workload shows any regressions. Recall that such savings are expected but \emph{opportunistic} (\cref{sec:prob}); i.e., since workloads are written against performance with HDD, our goal is to improve storage costs without degrading application performance relative to this baseline. Any additional performance savings are on top of these goals.

It should also be noted that the application-level performance change depends highly on application's workload composition, most notably the compute to I/O ratio. We select these applications for our evaluation because they are typical in the workloads we need to handle. Other applications' performance change could be very different in these scenarios. 

\subsection{Sensitivity Analysis}

\begin{table}[t]
\centering
\scalebox{0.85}{
\begin{tabular}{c|c|c}
Method & TCO Savings Percent & Model Top-1 Accuracy \\
\hline
Ours ($N=2$) & 9.25\% & 73.4\% \\
Ours ($N=5$) & 11.1\% & 55.6\% \\
Ours ($N=15$) & 12.7\% & 32.3\% \\
Ours ($N=25$) & 12.6\% & 24.2\% \\
Ours ($N=35$) & 10.8\% & 21.2\% \\
Best Baseline & 10.7\% & /
\end{tabular}
}
\caption{The TCO savings under different category numbers.}
\label{tab:tco-category-number}
\end{table}

\begin{figure}
    \centering
    \includegraphics[width=0.3\linewidth]{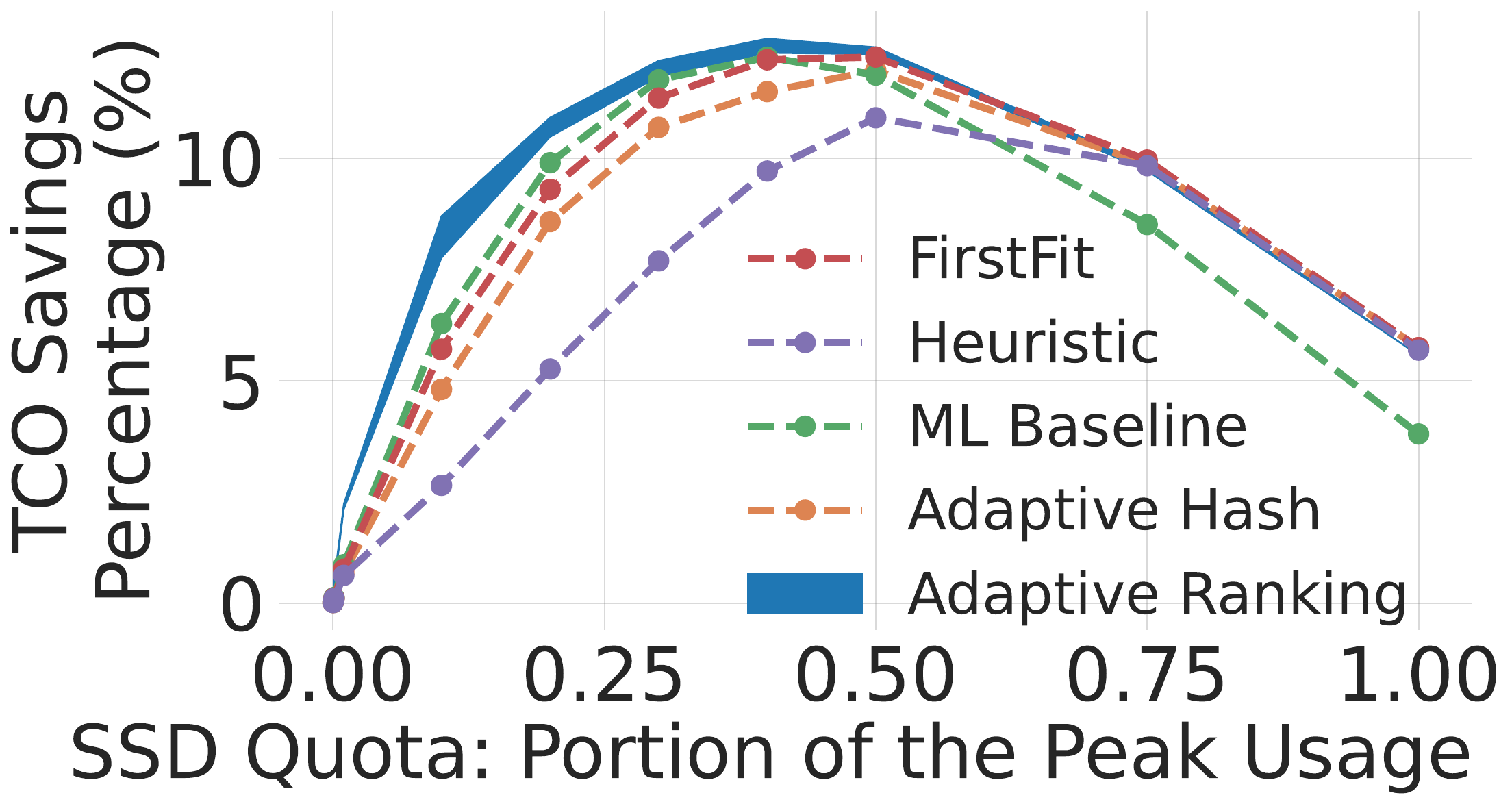}
    \caption{Adaptive algorithm parameters sensitivity.}
    \label{fig:tco-saving-hyperparameter}
\end{figure}

We explore the sensitivity of our method under different hyperparameters below.

\textbf{Adaptive Algorithm Parameters.} We include all combinations of hyperparameters where $T_{\textsc{SpilloverTCIO}} \in \{[0.005, $ $0.03],$ $[0.01,$ $0.15], [0.05, 0.25]\}$, look back window time length (seconds) $t_w \in \{600, 900, 1800\}$, and admission decision effective time $t_l \in \{600, 900,$ $1800\}$. We evaluate the sensitivity of the TCO savings for the same set of workloads in \cref{fig:tco-curve}. For each parameter combination, we apply the same parameter settings to all the workloads in the group. In \cref{fig:tco-saving-hyperparameter}, the blue area in the figure presents the upper bound and lower bound of TCO savings under different SSD capacities across different hyperparameter combinations. Our solution is not sensitive in terms of hyperparameter selection in the adaptive algorithm.

\textbf{Sensitivity on Category Numbers.} Our evaluation utilizes the $0.1$ SSD portion setting with all the algorithm parameters maintain consistent. It is critical to select an appropriately large number of categories to enable the model to effectively distinguish the cost across jobs without increasing the model's capacity for fine-grained category prediction. We present the impact of category numbers $N$ on end-to-end TCO savings in \cref{tab:tco-category-number}. A model with smaller category number achieves higher accuracy but fails to optimize the end-to-end TCO savings due to its limited granularity. Conversely, increasing the number of categories enhances granularity but at the cost of accuracy, diminishing the TCO savings.

\subsection{Adaptive Category Selection Dynamics}

\begin{figure}[t]
    \centering
    \begin{subfigure}{0.6\linewidth}
        \centering
        \includegraphics[width=0.98\linewidth]{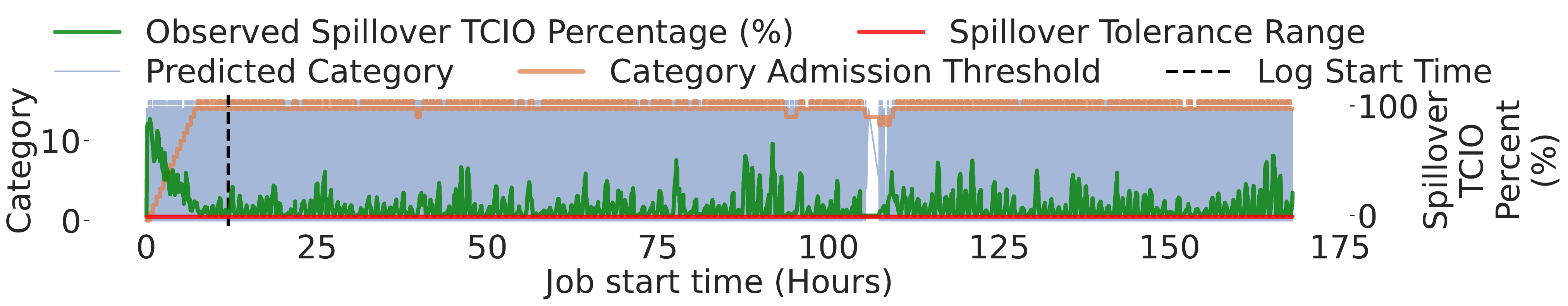}
    \end{subfigure}
    
    \begin{subfigure}{0.6\linewidth}
        \centering
        \includegraphics[width=0.98\linewidth]{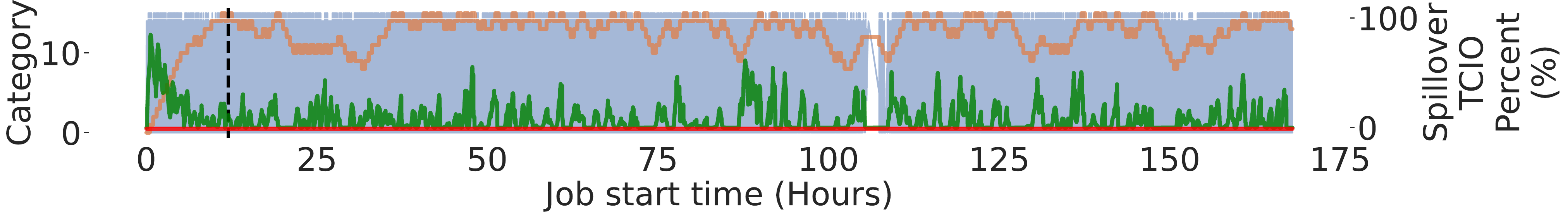}
    \end{subfigure}
    
    \begin{subfigure}{0.6\linewidth}
        \centering
        \includegraphics[width=0.98\linewidth]{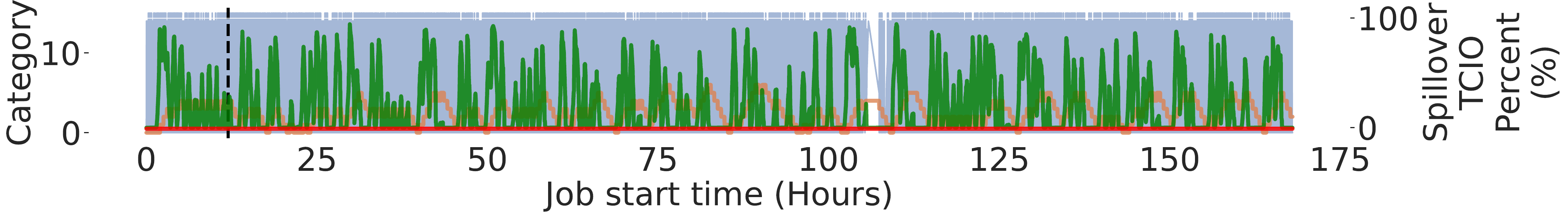}
    \end{subfigure}
    
    \begin{subfigure}{0.6\linewidth}
        \centering
        \includegraphics[width=0.98\linewidth]{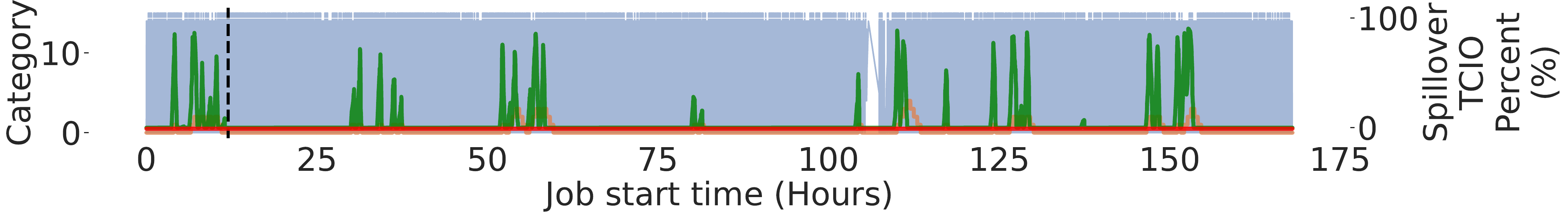}
    \end{subfigure}
    \caption{Category change of one workload. From top to bottom, the SSD quota covers 0.01\%, 1.0\%, 10\%, and 50\% of the peak SSD space usage under no SSD quota limit. The green line is the observed $\textsc{SpilloverTCIO}$ and the orange line represents the category admission threshold. The red area at the bottom of each figure is $T_\textsc{SpilloverTCIO}$.}\label{fig:category-pattern}
\end{figure}

To demonstrate the dynamics of our adaptive algorithm, we present the pattern of category threshold change and spill over percentage in \cref{fig:category-pattern}. We track the threshold change for 1 week online. Our adaptive category selection algorithm can adjust the category admission threshold to a higher range when SSD quota is limited and allow more category admissions when SSD space is plentiful.

\section{Detailed Related Works Discussion}
\label{app:related-works}

Prior works have shown the viability of machine learning for task property prediction in storage systems. \cite{hao2020linnos} leverages a small neural network to infer SSD performance with fine granularity and help parallel storage applications. The method learns a binary latency model and pre-calculate an inflection point for each model during a labelling stage. The key benefit is model simplicity and fine granularity of prediction, enabling more complicated applications online within latency requirements. \cite{zhou2021learning} tackles a problem related to our setting in data placement with methods that leverage application-level information and distributed traces in a way inspired by ideas from natural language processing. While the paper focuses on a specific learning problem of mapping textual metadata to storage-related properties, our work focuses on the practical designs and deployment of such models. 

Other applications of machine learning in storage systems include training one monolithic model for the entire storage system (not deployable in warehouse-scale setting due to adaptability): applying imitation learning for cache replacement to approximate an optimal oracle policy \cite{liu2020imitation}, guiding the placement algorithm model through reinforcement learning \cite{kaler2023deep, singh2022sibyl}; predicting properties in other aspects of data placement: improving a storage system through optimizing readahead and NFS read-size values with machine learning models \cite{akgun2023improving}, utilizing ML to improve on existing cache replacement strategies (LRU, LFU, etc.) \cite{vietri2018driving}, and predicting future task failures through ML \cite{chakraborttii2020improving}.

Multiple machine learning techniques have also been proposed in broader system problems \cite{kanakis2022machine, maas2020taxonomy}, ranging from resource allocation \cite{mishra2018caloree}, memory access prediction \cite{hashemi2018learning}, offline storage configuration recommendation \cite{klimovic2018selecta}, database query optimization \cite{kraska2021sagedb}, to networking applications \cite{dong2018pcc, abbasloo2020classic}. Although the nature of these applications is different from data placement in storage systems, they all show evidence that machine learning can be used in systems and benefits from domain-specific formulations.

\textbf{Data Placement in Practice.} Though machine learning for systems has been widely explored in different application domains, the state of the art practical solutions for caching or tiering in storage systems are still mostly heuristic.

Hadoop offers three caching schedulers: FIFO \cite{pakize2014comprehensive}, Capacity \cite{raj2012enhancement}, Fair \cite{zaharia2009job}. Spark supports FIFO, Fair. For each user, Azure tracks the last-accessed files and make the placement based of the self-tracked access history \cite{azureTiering}. \cite{yang2022cachesack} presents a novel adaptive cache admission solutions for Google, of which we implement a modified version in our comparison.

Very recent works have also started rethinking the best practical solution within the heuristic-based domain. \cite{yang2023fifo, yang2023fifo2} consider a modified FIFO for cache eviction, which achieves good scalability with high throughput on production traces from Twitter and MSR. \cite{eytan2020s} revisits the effectivenss of LRU versus FIFO and finds that FIFO exhibits better overall cost than LRU on production traces, including IBM COS traces. \cite{zhao2023tectonic} proposes new heuristics for storage, specifically tailored for machine learning workloads at Meta.

Another noteworthy work presents a solver-based solution for task scheduling in the setting where each task contains a list of preferred locations identified prior to scheduling. Their approach formulates the problem as a minimum cost maximum matching problem \cite{herodotou2021trident}. Although closely related to our work, as discussed in the \Cref{sec:bg} and \Cref{sec:prob}, the method is not directly feasible in our context. The primary challenge in adopting such a solver-based approach in our setting lies in the lack of jobs' cost at scheduling time.
\section{Discussion}
\label{app:discussion}

Our ``bring your own model'' idea can be adopted in other deployments and frameworks. The solution flow of our system is: 1) Setting an optimization objective (TCO savings in our case). 2) Designing a `hint’ (workload model output) passed from each workload from the application layer to the caching layer. Our hint is job importance in terms of TCO savings. 3) Picking jobs of different categories based on feedback from system utilization.

This BYOM flow remains adaptable to other applications. The key components that vary are the objective function, available model features, and how system utilization is quantified.

While \Cref{app:tab:detail-feature} lists specific features used in our model, these features fit into four general categories: historical information, job start time, execution information, and allocated resource information. The general categorization remains applicable though the detailed features may vary, allowing for adaptation across different organizations.

\end{document}